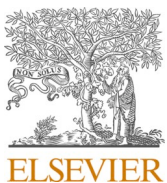

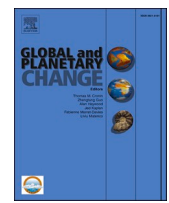

Invited Review Article

# Quantification and interpretation of the climate variability record

Anna S. von der Heydt [a,*], Peter Ashwin [b], Charles D. Camp [c], Michel Crucifix [d], Henk A. Dijkstra [a], Peter Ditlevsen [e], Timothy M. Lenton [f]

[a] Institute for Marine and Atmospheric research Utrecht, Department of Physics, Utrecht University, Princetonplein 5, 3584 CC Utrecht, The Netherlands
[b] Centre for Systems, Dynamics and Control, Department of Mathematics, University of Exeter, Exeter EX4 4QF, UK
[c] Mathematics Department, California Polytechnic State University, San Luis Obispo, CA 93407, USA
[d] Université catholique de Louvain, Earth and Life Institute (ELIC), Place Louis Pasteur 3, BE-1348 Louvain-la-Neuve, Belgium
[e] Centre for Ice and Climate, Niels Bohr Institute, University of Copenhagen, Denmark
[f] Global Systems Institute, University of Exeter, Exeter EX4 4QE, UK

ABSTRACT

The spectral view of variability is a compelling and adaptable tool for understanding variability of the climate. In Mitchell (1976) seminal paper, it was used to express, on one graph with log scales, a very wide range of climate variations from millions of years to days. The spectral approach is particularly useful for suggesting causal links between climate variability and climate response variability. However, a substantial degree of variability is intrinsic and the Earth system may respond to external forcing in a complex manner. There has been an enormous amount of work on understanding climate variability over the last decades. Hence in this paper, we address the question: Can we (after 40 years) update the Mitchell (1976) diagram and provide it with a better interpretation? By reviewing both the extended observations available for such a diagram and new methodological developments in the study of the interaction between internal and forced variability over a wide range of timescales, we give a positive answer to this question. In addition, we review alternative approaches to the spectral decomposition and pose some challenges for a more detailed quantification of climate variability.

## 1. Introduction

Imagine that the rapidly changing surface temperature at a fixed latitude and longitude, say every minute, was recorded throughout the entire history of the Earth. The temperature would fluctuate on a minute-by-minute basis as clouds pass in front of the Sun. The diurnal cycle of day and night would stand out. In the extra-tropical regions, the changing weather of highs and lows passing by would vary on a weekly scale, and the seasons would be apparent in the annual cycle. On the longer term, the El Niño Southern Oscillation (ENSO) would be imprinted as well as decadal to multidecadal variability in the oceans. Over the past million years, glacial cycles would dominate and on even longer, geological, timescales, tectonic processes (drifting continents) change ocean basins and flow patterns, while biological evolution affects the chemical balance of the atmosphere and oceans.

A compact way of presenting recurring variations of some observation over time scales is through a power spectrum. Such temperature spectrum is often referred to as the climate spectrum. This is the cornerstone of a paradigm of climate variability that appeared in the 1970s, in particular in the seminal paper of Mitchell (1976) and since then it has been a powerful organizing principle to understand climate variability (Ghil, 2002). The original "Mitchell spectrum" (MS) shown in Fig. 1 was based on a handful of observations, available in the mid 1970s, such as the central England temperature time series, the first ice-core and ocean sediment core records, and routine atmospheric data (U. S. Committee for the Global Atmospheric Research Program, 1975).

The MS indicates that climate on Earth varies on many temporal and spatial scales but that some time scales have more spectral density than others. The main source of energy available to the climate system is the incoming solar radiation (insolation). A typical spectrum of the surface temperature exhibits several peaks (of varying sharpness) that show up above a "continuous background" spectrum. The sharper peaks are more easily interpreted and tend to be associated with periodicities in external quasiperiodic[1] forcing related to the diurnal, yearly (seasonal) and longer periods caused by changes in the orbital configuration of the Earth, though not always unambiguously. For example, the peak at 40

---

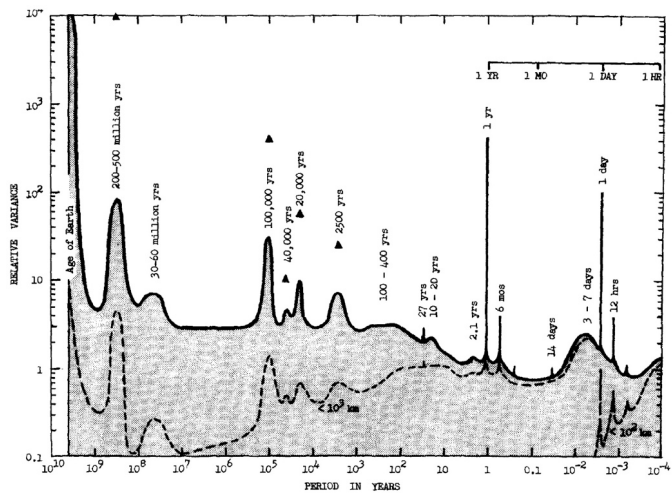

**Fig. 1.** An artists' view of climate variability (Mitchell, 1976) displaying a 'hypothetical' spectrum based on information of different time series. See Fig. 2 for an updated version.

ka is commonly interpreted as the signature of the response to changes in Earth's obliquity (see Section 2.4), but it may also be a period-doubling response to the quasi-periodic revolution of the longitude of the perihelion, of which the period is around 20 ka (Verbitsky and Crucifix, 2020). The broader peaks/bands in turn often involve physical phenomena within the climate system that vary on "typical" time scales but not on a clearly defined period; moreover, these variations are in several cases related to a specific spatial pattern. Finally, there is the continuous background suggestive of apparently random variability at lower frequencies. A climate spectrum will look somewhat different depending on which observable is used; the most common one is surface temperature - either globally averaged or at a specific location on Earth. Other relevant observables may be precipitation (location dependent), sea level or (sea) ice extent, however, these depend less directly on the radiative balance and, therefore, their response to external radiative forcing will be more difficult to interpret.

The meaning of the MS has been under quite some debate (Huybers and Curry, 2006; Ditlevsen et al., 1996). While the sharp peaks related to forcing periods are without doubt, the statistical significance of the other broader peaks cannot be assessed in many cases (in particular for the longer time scales). There are also characteristics of variability that are not well represented in a spectrum, including (abrupt) shifts in regime caused by transitions. Time series of geological records (which are mostly interpreted as "stacks" rather than individual time series) sometimes exhibit large and abrupt steps. For example in the deep sea oxygen isotope record of the Cenozoic period (65 million years) the glaciation of the Antarctic continent stands out as one large and rapid shift (Zachos et al., 2001). This transition involves typical time scales of the evolving processes such as the ice sheet growth, global ocean circulation and carbon cycle processes and would therefore contribute in some way to spectral bands related to ice and ocean variability. The spectral view cannot distinguish between large shifts and recurring events in the relevant climate subsystems and, maybe more importantly, says little about their predictability. Thus both the spectral and complementary representations of time series are necessary for a comprehensive description of the climate variability.

Furthermore, the origin and quantification of the continous background has remained unclear for a long time. The different paleoclimatic proxies give a consistent picture, showing that the continuous part of the climate spectrum accounts for a large part of the total variance (Lovejoy and Schertzer, 2013a). Many spectra of climate time series exhibit power law behavior, often related to scaling behaviour, which is a fundamental property of many physical and biological systems (Franzke et al., 2020). The background spectrum is fairly well described by "scaling regimes", where the spectral density scales with the frequency $f$, $p(f) \sim f^{-\beta}$, with $\beta$ significantly smaller than 2 (the characteristic of a red noise process). The existence of such scaling regimes is suggested from inspection of double logarithmic plots, i.e. showing the logarithm of the variance as a function of the logarithm of the period (as used in the MS, Figs. 1, 2) or frequency, and the scaling exponents $\beta$ can be estimated from the slopes of linear regressions to (parts of) the graph. It has also been proposed that "one-over-f" noise is a better description of the climate spectrum (Rypdal and Rypdal, 2016). The dynamics that would be responsible for such an observed spectrum is non-trivial. The generic example of dynamics leading to this kind of scaling in time scales is the cascade processes in rotating turbulent flow, which also dominates the climate on time scales of the atmosphere and oceans, but it cannot explain the climate spectrum on time scales much longer than the coherence times of the geophysical flows. More recently, long-memory processes in the heat equation are being discussed, which could potentially explain some of the longer time scale scaling regimes (Lovejoy, 2020; Fredriksen and Rypdal, 2017).

Over the past decades, there has been considerable progress in understanding climate variability by large observational efforts, model simulations and theory development. In this review we address the issue whether the MS diagram should be essentially updated and provided with a novel interpretation after 40 years of research. Such an update is important for at least two reasons: (i) to provide better estimates of the natural variability in the climate system, in particular on the time scales of present climate change (interannual-multidecadal time scales), and (ii) for the development of a theory of climate variability capturing the effects of the interaction of the intrinsic dynamics with the external forcing.

The aim of this paper is to present a modern view of climate variability, caused by internal processes, external forcing and their interaction, and discuss potential mechanisms of non-stationary forcing interaction with unforced processes working at a variety of time scales. In doing so we update Mitchell (1976) to include discussion of some of the more recent developments in climate variability. The spectral focus of this update is presented in Fig. 2. The "landscape" at the bottom of the figure represents the responses of the climate system departing from a 1/$f$ background spectrum, using colours that correspond to the various physical processes driving the variability. The top of the diagram indicates various sources and relative strengths of periodic (orbital) and random (impact/volcanic/solar) forcing variability. Between the landscape and the forcing we indicate the nomenclature of Lovejoy and Schertzer (2013a) for various timescale regimes. Table 1 lists some of the modes of variability represented in Fig. 2, together with references to sections in this paper with more details.

The remainder of the paper is organised around the updated Mitchell spectrum in Fig. 2. We start in Section 2 by describing the observational records (Section 2.1) and associated models (Section 2.2) that have become available since the MS. In Section 2.3 we discuss recent developments in determining the background spectrum, followed by detailed discussions of externally forced variability (Section 2.4), the modes of internal variability listed in Table 1 (Section 2.5) and long time scale variability (Section 2.6), which most often is a mix between externally forced response and internal time scales. We then focus in Sections 3 and 4 on alternative approaches to the spectral view, which have been used to understand the dynamics of climate variability and the background spectrum. In particular, Section 3 discusses developments in methods to extract and detect preferred time scales and patterns in data time series. Section 4 discusses some theoretical developments useful for understanding the sources of modes of variability. We return to update the MS in Section 5 and argue that our understanding of climate variability has substantially developed since publication of the Mitchell (1976) paper.

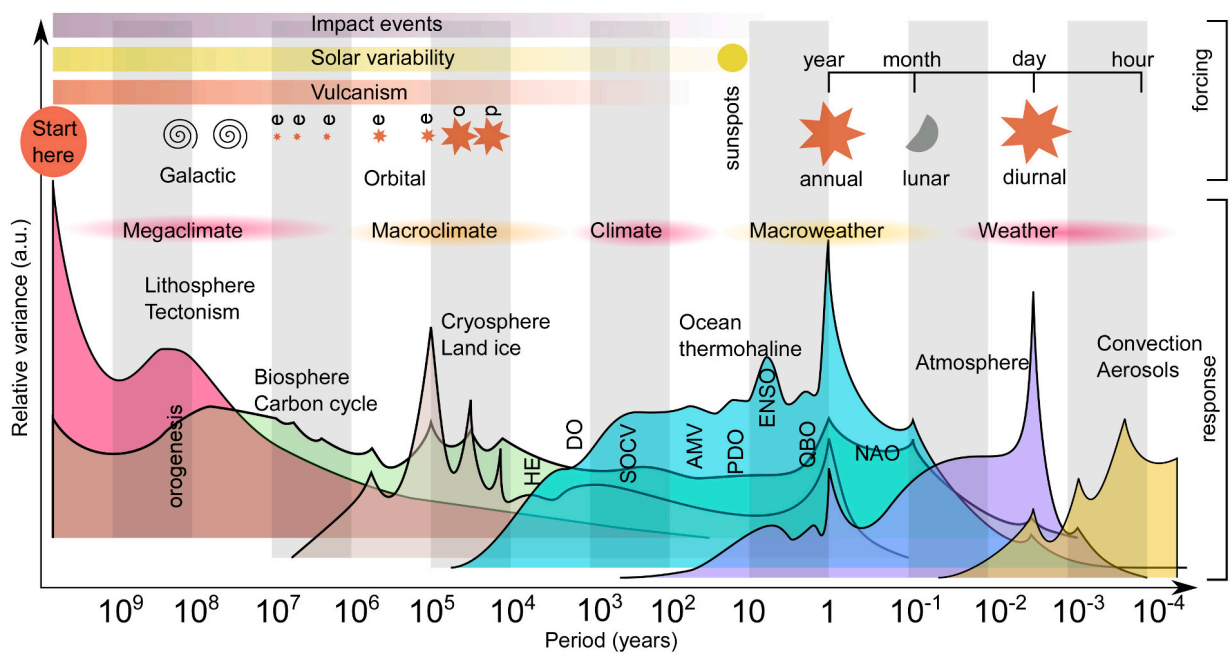

**Fig. 2.** A conceptual landscape of periodicities present in a typical climate signal from the atmosphere at the Earth's surface, updated after Mitchell (1976). The external forcing is shown in the upper part of the diagram, where approximate periodic forcing from solar system orbital behaviour. The landscape of responses and internal variability due to dynamics in various parts of the earth system are shown below, after removal of a background, where the stepped background indicates that short term variability is only known from recent records. Orbital variability due to eccentricity (e), obliquity (o) and precession (p) is shown, as well as timescales of galactic forcing. Impact events, solar variability and vulcanism events represent aperiodic forcing of highly variable amplitude. Known modes of variability with their approximate periodicities (broad peaks) are indicated with their acronyms (see Table 1) and explained in the text.

**Table 1**
Various modes of climate variability shown schematically in Fig. 2 that are discussed within the indicated sections of the text. The last two acronyms are abbreviations used in the text together with internal modes of variability, but are not explicit modes themselves.

| Mode | Name | Section |
|------|------|---------|
| AMV | Atlantic Multidecadal Variability | 2.5.3 |
| AMO | Atlantic Multidecadal Oscillation | 2.5.3 |
| DO | Dansgaard Oeschger Event | 2.5.4 |
| ENSO | El Niño-Southern Oscillation | 2.5.2 |
| HE | Heinrich Event | 2.5.4 |
| MJO | Madden Julian Oscillation | 2.5.1 |
| NAO | North Atlantic Oscillation | 2.5.2 |
| PDO | Pacific Decadal Oscillation | 2.5.2 |
| IPO | Interdecadal Pacific Oscillation | 2.5.2 |
| QBO | Quasi-biennial Oscillation | 2.5.1 |
| SOVC | Southern Ocean Centennial Variability | 2.5.3 |
| (A)MOC | (Atlantic) Meridional Overturning Circulation | 2.5.3 |
| MPT | Middle-Pleistocene Transition | 2.6.1 |

## 2. Types of variability - a journey through time scales

The daily and seasonal variability are directly related to processes external to the climate system and are therefore considered as forced variability. However, most of the variability arises through processes internal to the climate system, mainly because of the existence of positive feedbacks; such variability is called internal or intrinsic climate variability. Both the internal and forced variability are often referred as natural variability as this would exist without the presence of humans on Earth. Climate variability is also caused by emissions of greenhouse gases due to human activities; such variability is referred to as anthropogenic climate variability, or more commonly climate change.

### 2.1. New observations

Many new observations have become available since the MS, which have especially revealed new types of climate variability on longer time scales (from interannual to centennial/millennial). Since the 1980s satellite observations of the Earth's surface have become available providing a rather continuous and spatially resolved record of e.g. sea surface temperature, sea surface height and sea ice cover. These data are most interesting for shorter time scales such as seasonal to interannual variability and their spatial patterns. Moreover, available observations since about 1850 have been put together into gridded data sets of sea surface temperature and sea ice (HadISST, Rayner et al. (2003)) and atmospheric surface temperature over both land and ocean (HadCRUT, Morice et al. (2012)). These data sets have also been combined with proxy datasets, e.g. Mann et al. (2008). These approximately 150 year long data sets provide a wealth of information on temperature variability. Low frequency variability has been found in the observational climate data in the 1980-90s (e.g. Folland, 1986; Kushnir, 1994 for the North Atlantic). However, for these rather long (multi-deacdal) time scale variations the instrumental record is relatively limited (e.g. Caesar et al. (2018)).

For longer times series, we must rely on indirect proxy data to reconstruct climate variability. Also, since the MS has been published numerous historical and palaeo records as well as new proxies have become available. The PAGES 2-K initiative provides a coordinated effort to reconstruct climate globally over the last 2000 years and has revealed regional-scale multidecadal variability patterns and more recently also global-scale multidecadal climate variability (PAGES 2k Consortium, 2013, 2019). On even longer time scales reconstructions of global temperature since the Holocene (Marcott et al., 2013) reveal climate changes such as the Little ice age. Individual, millennial scale records of sea surface temperature at specific locations (Tasmanian summer temperature, Cook et al. (2000), sea temperature near Iceland, Sicre et al. (2008)) and at sufficient temporal resolution have revealed even longer, centennial time scale variations of temperature. Many efforts have been put into the reconstruction of global-scale sea ice variations on long time scales, e.g. de Vernal et al. (2013).

When establishing the climate spectrum from proxy data we have to take uncertainties into account. These fall in different categories: Firstly, a proxy is an indirect measure, where in most cases some transfer function relates the proxy and the climate variable of interest. Secondly, the measurement can be subject to measurement uncertainty or variations unrelated to climate and, thirdly, there might be uncertainty in dating the proxy. These three categories of uncertainty (and possibly more), influence the spectrum obtained from proxies in different ways. The transfer function between proxy and climate variable can be different at different temporal scales: A proxy like the tree-ring records will typically ignore low frequency variability, where climate adaptation over centuries masks very long term climate variations. Thus the transfer function acts like a high-pass filter, while in other proxies, like sediment records, mixing and diffusion will smooth out short term climate variations and the transfer function acts like a low-pass filter. Direct measurement uncertainty or independent uncorrelated noise in the proxy will typically add a white noise component to the spectrum. The third category, dating uncertainty in the proxy, will widen spectral peaks, i.e. move spectral density from peaks in the spectrum to the continuous background spectrum (Ditlevsen et al., 2020).

## 2.2. The climate model hierarchy

Observations are crucial to study climate variability, but the records remain very limited, in particular for the longer time scales. As we also cannot investigate climate variability phenomena in the laboratory, climate models have become a central tool in climate research, in particular for understanding the processes that cause the observed climate variability. A wide range of model types is in use, from low-order dynamical systems to high resolution Global Climate Models (GCMs).

Scales and processes are important properties of climate variability and this motivates to classify climate models using two traits (Fig. 3). Here the trait "scales" refers to both spatial and temporal scales as there exists a relation between both: on smaller spatial scales usually faster processes take place. "Processes" refers to either physical, chemical or biological processes taking place in the different climate subsystems (atmosphere, ocean, cryosphere, biosphere, lithosphere).

Models with a limited number of processes and/or scales are usually referred to as conceptual climate models. In these models only very specific interactions in the climate system are described, while their state-vector can still be large, e.g. covering a large range of scales for a few climate variables. An example are models of glacial-interglacial cycles (Saltzmann, 2001) formulated by small-dimensional systems of ordinary differential equations. For example, only the interactions of the ice sheet volume, atmospheric $CO_2$ concentration and global mean ocean temperature is included.

Keeping the number of processes limited, more scales can be added by discretizing the governing partial differential equations spatially up to three dimensions and moving upwards in the diagram to *Intermediate Complexity Models* (ICMs). A higher spatial resolution and inclusion of more processes will give models located in the right upper part of the diagram. In a GCM, the atmosphere, ocean, ice and land components are divided into grid boxes. Over such a grid box the budgets of momentum, mass and for example heat are considered. The state-of-the-art GCMs are located above the Earth System Models of Intermediate Complexity (EMICs, Claussen et al. (2002)) because they represent a larger number of scales. Compared to GCMs, the ocean and atmosphere models in EMICs are strongly reduced in the number of scales. For example, the atmospheric model may consist of a quasi-geostrophic or shallow-water model and the ocean component may be a zonally averaged model. The advantage of EMICs is therefore that they are computationally less demanding than GCMs and hence many more long-time scale processes, such as land-ice and carbon cycle processes can be included. Each of the individual component models of EMICs may also be used to study the interaction of a limited number of processes, ending up in the ICM category. A prominent example is the Zebiak-Cane (ZC) model of the El Niño/Southern Oscillation phenomenon (Zebiak and Cane, 1987). In time, the GCMs of today will be the EMICs of the future and the state-of-the-art GCMs will shift towards the upper right corner in Fig. 3.

Regarding the MS, the models can give two contributions: (i) whether certain preferred time scales of variability exist and (ii) provide specific mechanisms of variability. The GCMs are better suited for issue (i) as they include a multitude of processes and scales. On the other hand, the conceptual models are better suited for issue (ii), i.e. mechanism identification. Models can also give insight into the different spectral scaling regimes identified in observations (Lovejoy et al., 2013). When one starts with a model high up in the hierarchy it may be difficult to connect the description of causal chains to well understood 'building blocks' as present in low-order dynamical systems (conceptual models). In this case, there must be a commitment to in-depth analysis of the model results. When starting with a model low in the hierarchy the connection to observations will be difficult. Hence, there must be a commitment to demonstrate that the processes identified in the conceptual model are also dominant in models higher up in the hierarchy. Theories then develop in stages and towards the interpretation of the observations using successively better models of which the behavior is better understood.

This approach will be used in Subsection 2.5 on internal climate variability below. We use results from global climate models (GCMs) from the CMIP5 archive, which have been used in the 5th assessment report of the Intergovernmental Panel on Climate Change (IPCC). The CMIP5 archive contains three types of model simulations:

(PI) Long (multi-centennial) control simulations under constant (pre-industrial) forcing (solar, greenhouse gases, aerosols).

(HIST) Simulations under historical forcing conditions, usually over the period 1850-2010. An end point of the PI simulation provides the initial conditions and multiple realizations (under slightly perturbed initial conditions) have been performed.

(RCP) About 100 year simulations, starting in 2010 and continuing up to the year 2100 under a Representative Concentration Pathway (RCP) scenario, as determined from Integrated Assessment Models (IAMs).

Regarding the detection of preferred variability, the PI simulations are most relevant since the forcing is constant except for the seasonal cycle. Hence, if preferred climate variability is detected in these models on interannual-to-centennial time scales it must be intrinsic under these conditions. Also, spectral scaling regimes (weather, macro-weather and climate) can be identified from these simulations by multifractal analysis (see next Subsection 2.3, Lovejoy and Schertzer (2013b)).

The HIST simulations are crucial to assess the quality of a particular model as these results can in principle be compared to observations; external forced changes such as long-term trends or effects of volcanic forcing can be directly compared to observations, while internal modes of variability can be compared only in a statistical sense, e.g. their amplitude, time scale and spatial pattern. However, as there is a time-dependent forcing in these simulations (e.g., changes in greenhouse gas concentrations, volcanic activity and solar irradiance), the results show the interaction of intrinsic variability with forced changes. The same holds for the RCP simulations which deal with projected forcing conditions.

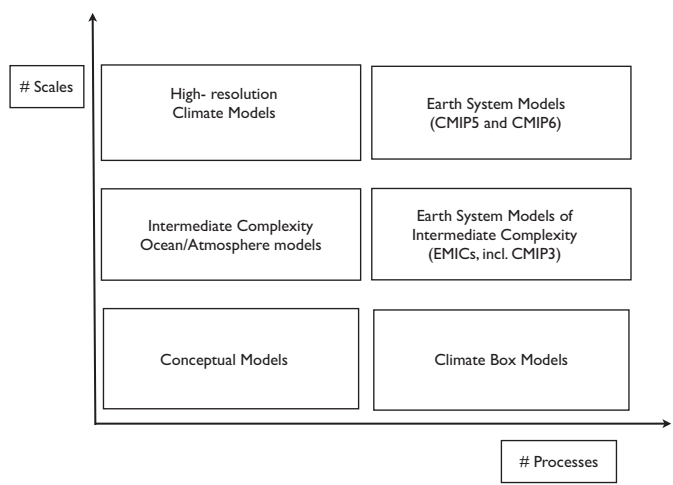

**Fig. 3.** Classification of climate models according to the two model traits: number of processes and number of scales. There is of course overlapping between the different model types, but for simplicity they are sketched here as non-overlapping.

An overview of the PI and HIST simulations can be found in Cheung et al. (2017) where also the number of historical ensembles and length of the control simulation is shown. These simulations mostly cover over 500 years and hence interannual-to-centennial variability due to intrinsic processes can be analysed in these simulations.

## 2.3. The background spectrum

Climate variability covers a huge range of spatial and temporal scales. Other than preferred frequencies (and spatial scales) of temperature variability the spectrum provides insight into the dynamics of the Earth system as a whole. Power-law behaviour, often related to scaling behaviour, in climate time series is quite frequently observed (Franzke et al., 2020). The most prominent example is the scaling behaviour of the energy cascade in homogeneous turbulence, which has no preferred scale, but provides insight into the dynamical interaction of different scales. On top of the 'background spectrum', which may exhibit scaling behavior, preferred frequencies or spatial patterns may be distinguished - set by specific modes of climate variability.

With the availability of long and high-resolution datasets, in particular ice core data (EPICA Community Members, 2006), it became possible to study scaling properties of these time series. These new analysis methods (see Section 3) aim to determine how fluctuation levels at each time scale are related.

To illustrate the scaling idea, consider a self-similar stochastic process $X(t)$, with $\lambda$ a scale factor, it holds that

$$X(\lambda t) = \lambda^H X(t) \tag{1}$$

where the equality holds in the distribution sense and $H$ is the Hurst exponent. One can show that the spectral power of such a process, $S(f)$, is given by a power-law distribution

$$S(f) = S_0 f^{-\beta} \tag{2}$$

where $\beta = 2H + 1$. Hence, there is a specific relation between fluctuations at different frequencies, given by the power law exponent $\beta$.

To determine scaling behavior in a time series, say $T(t)$, one considers the behavior of fluctuations $\Delta T(\Delta t)$ versus $\Delta t$, $\Delta t$ being a multiple of the minimum time scale. The fluctuations are calculated either by wavelet transform coefficients or by a simple Haar wavelet difference,

$$\Delta T(\Delta t) = \frac{2}{\Delta t} \left| \sum_{t}^{t+\Delta t/2} \widetilde{T}(t) - \sum_{t+\Delta t/2}^{t+\Delta t} \widetilde{T}(t) \right| \tag{3}$$

where $\widetilde{T}(t) = T(t) - \overline{T}$ and $\overline{T}$ the mean of the time series. The fluctuations display scaling behavior, when the fluctuation function

$$S_q(\Delta t) = \langle (\Delta T(\Delta t))^q \rangle \sim (\Delta t)^{\zeta(q)} \tag{4}$$

where the brackets indicate ensemble mean and $\zeta(q) = qH - K(q)$, where $K(q)$ is an anomalous scaling factor often attributed to intermittency.

In most studies, $q = 2$ is taken and the fluctuation function $S(\Delta t)$ is defined as

$$S(\Delta t) = S_2(\Delta t)^{\frac{1}{2}} \sim (\Delta t)^{(2H - K(2))} \tag{5}$$

As the intermittency measured by $K(2)$ can be shown to be small, this indicates that $S(\Delta t) \sim \Delta t^H$. When $H < 0$ and hence $\beta = 2H + 1 < 1$, fluctuations will decrease with scale $\Delta t$ and when $H > 0$, and $\beta > 1$ fluctuations will increase will scale $\Delta t$.

An analysis of a wide range of observational time series has led to the identification of different scaling regimes. Up to about 5-10 days, a power-law spectrum with $\beta \approx 2$ is found, giving $H \approx 0.5$, which is the Hurst exponent of Brownian motion. Up to about 50 years, the macro-weather regime is characterized by $\beta \approx 0.2$, giving $H \approx -0.4$. In the climate regime, up to 50 kyr, a positive Hurst exponent $H \approx 0.2$ is found, associated with $\beta \approx 1.4$ (e.g. Lovejoy et al. (2013)). Other regimes at even larger time scales have also been identified (Lovejoy, 2018). The analysis of output of four GCM control simulations (Lovejoy et al., 2013) indicates that only the macro-weather regime is simulated by these models.

One interpretation of these results is that apart form phenomena directly attributable to an external forcing, such as the annual and daily cycle and those associated with volcanic eruptions, all climate variability is contained into this background signal (Lovejoy and Schertzer, 2013b). The background signal originates from multifractal cascading processes, such as those in turbulent flows. While it is widely accepted that these processes lead to fluctuation variations in the weather regime, other non-multifractal processes may be responsible for the scaling behavior at longer time scales. Moreover, variability associated with specific phenomena, such as the El Nino-Southern Oscillation or Dansgaard-Oeschger events events, may not be part of this scaling regime.

For example, Rypdal and Rypdal (2016) show that when the scaling analysis is applied to interstadial and stadial ice core data separately (hence no DO events are included), the scaling exponent $\beta \approx 1$, similar to the macroweather regime. Hence, they suggest that the so-called 1/f noise ($\beta = 1$) scaling holds over both the macro-weather and climate regimes. High-resolution proxy data have substantially contributed to a better estimation of these Hurst exponents (Veizer et al., 1999; Huybers and Curry, 2006).

## 2.4. Externally forced climate variability

Several external sources of climate variability are identified in Mitchell (1976) where he divides them in (a) deterministic (i.e. quasi-periodic) solar and lunar forcing (b) other deterministic forcing and (c) volcanic activity. Forcing under (a) includes not only the obvious diurnal and annual forcing but also long-timescale *astronomical forcing* changes, summarised in e.g Crucifix et al. (2006). The latter consists of changes of the distribution of insolation resulting from the quasi-periodic changes in the parameters determining the orbit of the Earth around the Sun and its obliquity, and is therefore also called the *orbital or astronomical forcing* of the climate system.

The amount of insolation energy available at the top of the atmosphere at a given time of the year is, to first order, a function of the total amount of energy received from the Sun at the mean Earth-Sun distance, the position of the Earth on its orbit (measured from a point of reference which determines the start of the cycles of seasons called the vernal equinox) and the angle of inclination (obliquity $\varepsilon$) of the Earth's equator on the orbital plane. The Earth's orbit is eccentric, where the eccentricity $e$ varies on long time-scales.

The revolution of the Earth around the Sun generates the seasonal forcing which, with the daily rotation, constitutes certainly the most prominent spectral peak of most climatic variables. Compared with most typical time scales of climate variability that we consider here, the seasonal cycle is clearly on a rather short time scale. Nevertheless, it is at least strongly interacting with many slower modes of variability; for example, interannual phenomena such as the El Niño-Southern Oscillation (ENSO) tend to be locked to the seasonal cycle (Rasmusson and Carpenter, 1982).

Changes of the distribution of insolation resulting from the quasi-periodic changes in obliquity, eccentricity and the longitude of the perihelion generate the orbital forcing of the climate system. Considerations on the symmetry make it clear that annual mean insolation does not depend on obliquity changes, nor on the longitude of the perihelion. Only the eccentricity component has an effect on the global averaged energy received by the sun. This can also be seen in Fig. 4a where the typical eccentricity time scales appear in a wavelet analysis. It varies following a complicated combination of periods, whose main components are around 100 and 400 kyr (Berger, 1978b; Laskar et al., 2004). An increase in obliquity decreases the annual mean insolation difference between the equator and the poles. It also increases summer insolation in both hemispheres. The revolution of the longitude of perihelion (called *climatic precession*) affects the insolation received by the globe at a given month. It therefore affects anti-symmetrically the seasonal contrast on both hemispheres, since summer on one hemisphere corresponds to winter on the other hemisphere. Therefore their typical time scales appear only when insolation is measured at a particular latitude see Fig. 4b). The global, annual mean amount of energy received varies according to a factor $(1 + e^2/2)$, i.e. of the order of 0.1%, which is usually considered as negligible. By contrast the changes in seasonal and meridional distributions have numerous effects on the dynamics of climate: from the mass balance of ice sheets, surface water temperatures, atmospheric circulation and precipitation regimes and consequently, ocean circulation, vegetation, weathering, to carbon and nutrient cycling.

Because of the quasiperiodic nature of much of celestial mechanics, the orbital forcing can be well approximated as a sum of a small number of sines and cosines. Berger (1978b) and Berger and Loutre (1991) give tables of amplitudes, frequencies and phases to this end. More accurate numerical computations of astronomical elements have since been made available (Laskar et al., 2011), but for a qualitative understanding of ice age dynamics the Berger decomposition remains particularly useful. The more accurate numerical solutions for the astronomical parameters have shown that due to the chaotic nature of the solar system they can only be calculated for the past approximately 50 million years, and not for longer time spans (Laskar et al., 2004).

Independent of the changes to the Earth's orbit there are several processes of solar variability that affect the insolation arriving at the Earth (Solanki, 2002). Best understood is the short-term variability that manifests as an oscillation with a period of 11 *yr* in sunspot number and an associated fluctuation in solar output - this is thought to be associated with a solar magnetic cycle driven by magnetoconvection within the solar core. Although the peak-to-trough variability in output is of the order of 0.1% which would imply little effect on the insolation forcing, there is evidence that the fluctuations in solar wind associated lead to atmospheric variability through stratosphere-troposphere coupling (Tomassini et al., 2011; Lu et al., 2011). Mitchell notes clear evidence of variability in solar activity and the 11 year sunspot cycle, but in the meantime there remains no clear evidence of this periodicity in the climate record; while at solar maxima the upper stratosphere warms and contains more ozone (Haigh, 1996), the climatic impact on e.g. atmospheric variability such as the North Atlantic Oscillation seems to be rather limited, (Chiodo et al., 2019). Moreover, Mitchell is equivocal about evidence of a response at 27 years from ''Brier''forcing; this was based on an apparent relation between global-scale pressure distributions and long-term lunar/solar tidal perturbations - in the meantime this has been dropped through lack of evidence.

Other deterministic forcing that Mitchell discusses under (b) includes orogenesis through speculated tectonic changes and/or passage through galactic dust bands. In the meantime, the science of tectonic changes is much better established, the galactic variability still speculative.

The plate tectonic motion that gives gradual remodelling of ocean basins over Myr timescales is clearly a forcing to the ocean-atmosphere system that will change circulation patterns, transport of energy and climate at a global level - for example, Zachos et al. (2001) link a number of changes in global climate epoch over the past 65 Myr to changes in ocean circulation associated with tectonic such as closing of the Panama seaway and the opening of the Drake passage. It has been suggested that the global cooling trend over the last 50 million years is at least in part due to continental movement as well as declining $CO_2$ (Sijp et al., 2014).

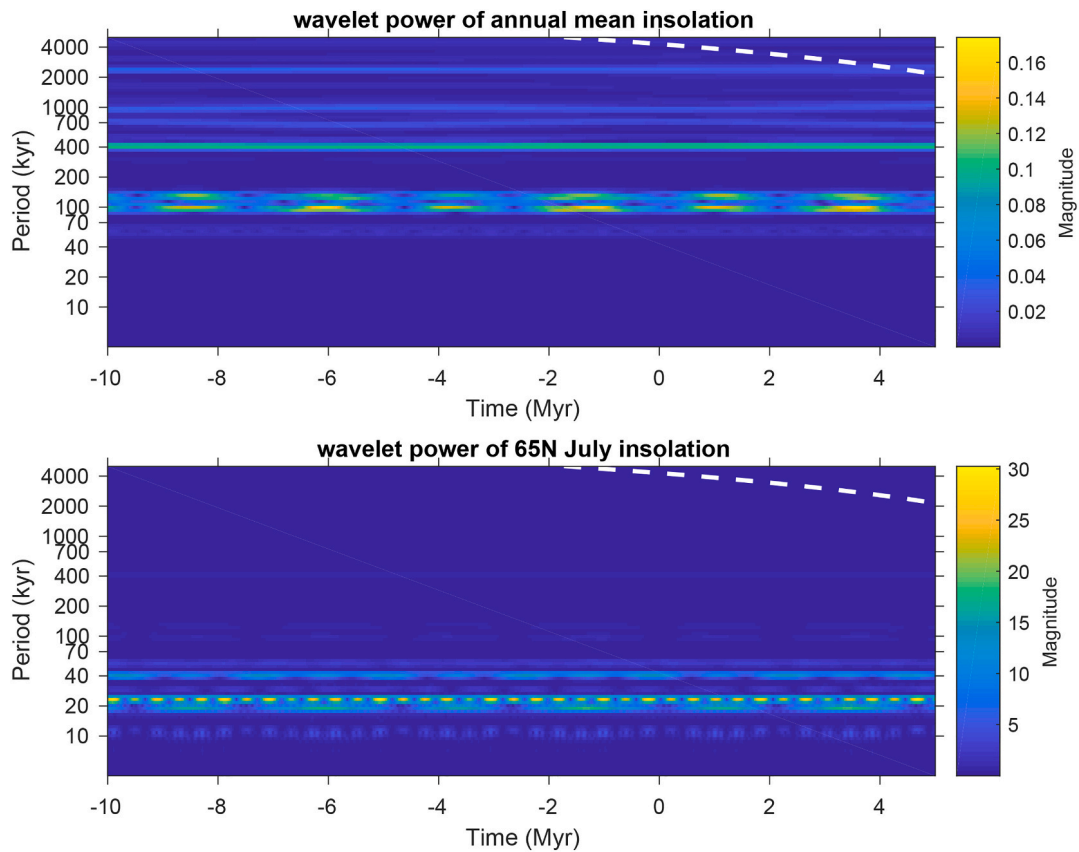

**Fig. 4.** Wavelet spectrum of the insolation on Earth calculated from the astronomical solution Laskar et al. (2004) for the period from 10 Myr ago until 5 Myr into the future. The analysis has been performed on a time series extending from 50 Myr in the past till 10 Myr in the future; white dashed lines indicate the wavelet cone of influence. (a) Annual and global mean insolation; (b) monthly mean (20 June - 21 July) insolation at 65'N.

A variety of other sources of external forcing have been suggested, notably changes in the position of the solar system within the galaxy. In particular, it has been suggested that the solar system passes between more and less dense stellar regions (as it passes through spiral arm regions) with a period of about 143 Myr, while the vertical oscillations of the solar system with respect to the galactic midplane are thought to occur every 64 Myr (Sloan and Wolfendale, 2013). This may in principle modulate cosmic ray background radiation which in turn may affect the climate via changes to biosphere and atmosphere, though Sloan and Wolfendale (2013) suggest the cosmic ray background effect may be not so large, and there is no clear signal on the expected frequencies.

On very long time scales, solar evolution over its lifetime changes solar output during geological time very slowly, suggesting that at the beginning of the Earth's life, the solar luminosity was about 25-30% lower than today (Solanki, 2002). This has produced the still unsettled *faint young Sun problem* for the early Earth's climate because the Sun provided insufficient energy to prevent the Earth from being completely ice-covered (Feulner, 2012).

Mitchell sees volcanic variability (c) as a modulation that involves climatic cooling for a few years. Changes in aerosol content and atmospheric composition associated with major volcanic activity will clearly affect the transmission of solar output into the atmosphere, and major eruptions (such as Pinatubo 1991) produce measurable climatic changes on the scale of several years. For example, McGregor et al. (2015) suggest that a cooling in the later half of the pre-industrial era 1-1800CE can be attributed to an increased frequency of volcanic eruptions. In the meantime, there is now a more established link between tectonic changes and climate (Lenardic et al., 2016) that may involve a wide variety of changes to the earth system, including volcanic activity, glaciation and surface weathering as well as ocean geometry.

In general, the climate system responds to external forcings providing variable energy input into the system. The result is a variable climate on many different time scales, as has been already suggested by the MS. However, it would be unreasonable to expect ice sheets (or other phenomena, such as the monsoons) to have an entirely linear response to the external (astronomical) forcing for two reasons: (i) A positive insolation anomaly may not have symmetrical effects to a negative insolation anomaly. (ii) The response of the climate system to the (variable) energy input will in general depend on the system state, in which case a form of multiplicative forcing occurs, which is a potentially strong factor of system instability and a potential route to chaos. The nonlinear response to forcing will be further discussed in Section 4.

### 2.5. Internal climate variability

In this section, we discuss phenomena of internal (or intrinsic) variability from shorter to longer time scales.

#### 2.5.1. Intraseasonal to interannual variability

From outgoing longwave radiation (OLR) anomalies, it was discovered in the early 1970s (Madden and Julian, 1972, 1994) that there is a strong eastward traveling signal in the tropics associated with anomalous rainfall. When there is strong precipitation (strong convection) the outgoing longwave radiation is lower (lower temperatures). The anomalies in precipitation are tightly coupled with the large-scale wind field. The eastward speed of the anomalies is about 4-8 m/s and hence the pattern travels from the west Indian Ocean to the east Pacific in about 30-60 days. It is called the Madden Julian Oscillation (MJO) but also referred to as the 30- to 60-day oscillation or simply intraseasonal oscillation. The occurrence of the MJO influences many tropical weather and climate phenomena (Kessler et al., 1996).

The most prominent mode of internal variability on inter-annual time scales is the El Niño-Southern Oscillation (ENSO). About once every 3-7 years, the sea surface temperature in the equatorial eastern Pacific is a few degrees warmer than normal (Philander, 1990). During the last decades, the equatorial Pacific has been observed in unprecedented detail (Timmermann et al., 2018) measuring relevant quantities in the equatorial ocean and atmosphere system, e.g sea-level pressure, sea-surface temperature, sea-level height, surface wind and ocean subsurface temperature. These observations were not available at the time the MS has been constructed, possibly explaining why the MS does not contain a specific peak at ENSO frequencies. Today we know that ENSO is an important phenomenon of natural (and internal) climate variability that has both a well-defined spatial pattern as a relatively well-defined time scale.

A measure of the state of ENSO is the NINO3.4 index, which is the area-averaged Sea Surface Temperature (SST) anomaly (i.e. deviation with respect to the seasonal cycle) over the region 170°W-120°W × 5°S-5°N. El Niño events typically peak in boreal winter, with an irregular period between two and seven years, and strength varying irregularly on decadal time scales. The spectral density of NINO3.4 at interannual time scales is clearly above a red noise spectrum, therefore adding to a modification of the MS. The spatial pattern of ENSO variability is often represented by methods from principal component analysis (Preisendorfer, 1988), detecting patterns of maximal variance. The first Empirical Orthogonal Function (EOF) of observed SST anomalies shows a pattern strongly confined to the equatorial region with largest amplitudes in the eastern Pacific.

The ENSO phenomenon is thought to be an internal mode of the coupled equatorial ocean-atmosphere system which can be self-sustained or excited by random noise (Fedorov et al., 2003). The mechanism of the ENSO mode growth and propagation is best captured by the recharge-oscillator view (Jin, 1997a), see Fig. 5. The interactions of the internal mode and the external seasonal forcing can lead to chaotic behavior through nonlinear resonances (Tziperman et al., 1994; Jin et al., 1994). On the other hand, the dynamical behavior can be strongly influenced by noise, in particular westerly wind bursts (Lian et al., 2014).

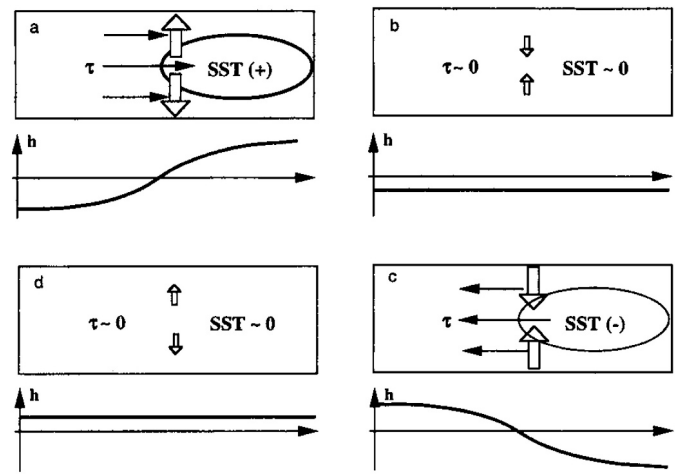

**Fig. 5.** The mechanism of the recharge-oscillator (Jin, 1997b). Consider a positive SST anomaly in the eastern part of the basin which induces a westerly wind stress ($\tau$) response (a). Through ocean adjustment, the slope in the thermocline (h) is changed giving a deeper eastern thermocline. Hence, through background upwelling the SST anomaly is amplified which brings the oscillation to the extreme warm phase (b). Because of ocean adjustment, a nonzero divergence of the zonally integrated mass transport occurs and part of the equatorial heat content is moved to off-equatorial regions. This exchange causes the equatorial thermocline to flatten and reduces the eastern temperature anomaly and consequently the wind stress anomaly vanishes. Eventually a nonzero negative thermocline anomaly is generated, which allows cold water to get into the surface layer by the background upwelling. This causes a negative SST anomaly leading through amplification to the cold phase of the cycle (c). Through adjustment, the equatorial heat content is recharged (again the zonally integrated mass transport is nonzero) and leads to a transition phase with a positive zonally integrated equatorial thermocline anomaly (d).

### 2.5.2. Decadal-to-multidecadal climate variability

On the decadal-to-multidecadal time scale, two modes of variability are most pronounced through their expression in sea surface temperature (SST), the Atlantic Multidecadal Variability (AMV), which in earlier studies has been termed Atlantic Multidecadal Oscillation (AMO), and the Pacific Decadal Oscillation (PDO), sometimes also referred to as Interdecadal Pacific Oscillation (IPO). The first analyses of North Atlantic variability (Schlesinger and Ramankutty, 1994; Kushnir, 1994) were based on observed sea-surface temperature (SST) and indicated the existence of variability on a time scale of 50-70 years. Warm periods were in the 1940s and from 1995 up to the present, whereas during the 1970s the North Atlantic was relatively cold. There is a negative SST anomaly near the coast of Newfoundland and a positive SST anomaly over the rest of the North Atlantic basin (Kushnir, 1994). Low-frequency variability in the North Atlantic SST has been determined from proxy data stretching back at least 300 years (Delworth and Mann, 2000) and within this data there is a statistically significant peak above a red-noise background at about 50 years. From recent Greenland ice-core analysis, where five overlapping records between the years 1303 and 1961 are available with annual resolution, significant multidecadal peaks in the spectrum were found (Chylek et al., 2011). Historical observations (since 1850) also suggest a significant spectral peak around 40-60 years (Mann et al., 2020).

Analysis of monthly mean SST data for the North Pacific (20-70°N) showed that the principal component (PC) of the leading EOF (Mantua et al., 1997) displays multidecadal variability. Subsequent analyses have indicated that the SST field has robust multidecadal statistical modes, where the first EOF is usually referred to the Pacific Decadal Oscillation (PDO). The second EOF is usually called the North Pacific Gyre Oscillation (DiLorenzo et al., 2008) and the PC also displays a multidecadal signal. More recently, longer records of historical observations, however, find no evidence for a significant PDO peak Mann et al. (2020).

From the CMIP5 PI simulations, there is clear evidence that the AMO/AMV index (defined as the SST anomaly over the North Atlantic, 0 - 60N, region) displays multi-decadal variability that extends above the background (red) noise (Han et al., 2016). The six models in Han et al. (2016) that show the 'best' agreement (in terms of correlation, amplitude and spatial pattern) indicate a dominant period of 20-70 years. Of these models, the GFDL-CM3 model is able to simulate the negative SST anomalies near Newfoundland, a feature found in observations, but difficult to capture in models. This model actually displays two significant periods of variability, one around 50 years and one around 25 years. On the other hand, Multi-taper Method Singular Value Decomposition (MTM-SVD) analysis of global SST fields in the CMIP5 control simulations reveals no significant multidecadal peak in the multi-model mean, while some few models do show AMV-like variability (Mann et al., 2020). Moreover, comparison between Atlantic multidecadal variability in CMIP5 control (PI) and historical (HIST) simulations suggests that a considerable part of the variability of the last 150 years can be explained by external forcing (Murphy et al., 2017), where aerosol forcing has been identified as potential candidate (Booth et al., 2012).

Cheung et al. (2017) investigate intrinsic variability both from the PI and HIST simulations, where they filter the forced signal in the latter simulations by using the multi-model ensemble mean. Their results show that the models clearly underestimate the intrinsic variability, both in the North Atlantic and in the North Pacific. At higher spatial resolution in one GCM it appears that the amplitude of intrinsic variability on multidecadal timescales increases, presumably because the mechanisms for this type of variability involves mesoscale features and instabilities (Jüling et al., 2020).

Inter-hemispheric effects between the Atlantic and Pacific (Dima and Lohmann, 2007) have been suggested as a mechanism of the AMO. Also conceptual models have identified several mechanisms which could be responsible for this preferred variability. An example is the 'thermal Rossby mode' mechanism (Fig. 6) as presented in Frankcombe et al. (2010). Here, westward propagation of temperature anomalies cause an out of phase response of the Atlantic meridional and zonal overturning circulation responses. Most suggested mechanisms for the AMO involve the ocean circulation, while a study with prescribed ocean heat transport also showed multidecadal Atlantic variability suggesting the AMO as response to stochastic atmospheric forcing (Clement et al., 2015). Indeed, it has been found that the thermal Rossby mode can be excited by stochastic noise with a spatial pattern resembling the North Atlantic Oscillation (NAO) Frankcombe et al. (2009).

Spectra for the CMIP5 PI and HIST simulations were summarized in the review by Newman et al. (2016). Although the background was generated by a linear inverse (multivariate AR(1)) model, there is no sign that the PDO displays variability beyond a red noise background (Mann et al., 2020). A sketch of the multitude of equatorial and midlatitude processes considered to be involved in the PDO is shown in Fig. 7. Oceanic Rossby waves, atmospheric noise, shifts in the Kuroshio Current and reemergence of SST anomalies due to ENSO are all considered to play a role (Newman et al., 2016).

### 2.5.3. Centennial-scale climate variability

Millennial scale records of sea surface temperature at specific locations (Tasmanian summer temperature, Cook et al. (2000), sea temperature near Iceland, Sicre et al. (2008)) has revealed even longer, centennial time scale variations of temperature. GCM simulations, however, studying such longer time-scale internal variability become scarce due to computational limitations. Here we discuss two examples of single GCMs having performed a few 1000 years of simulation.

The first example comes from a 4,000 year simulation carried out with the GFDL CM2.1 model under constant preindustrial forcing (Delworth and Zeng, 2012). The observable chosen was the surface air temperature averaged over the Atlantic domain and over the latitudes 20°N-90°N. This quantity shows dominant variability on a few hundred year time scale, for which the red noise null-hypothesis can be rejected. Careful analysis indicates that the inter-hemispheric heat transport associated with variations in the Atlantic meridional overturning circulation (AMOC) in the Atlantic is responsible for this variability. These variations arise through the advection of salinity anomalies by the AMOC, which also determines the time scale.

Can this variability be traced back to an amplification of a single pattern in a more idealized model? Indeed, while investigating instabilities of the AMOC in an idealized North Atlantic basin, it was found that buoyancy anomalies which propagate over the overturning loop can be amplified (Winton and Sarachik, 1993; Te Raa and Dijkstra, 2003; Sevellec et al., 2006); such oscillations are called 'Loop Oscillations' (or overturning oscillations). The mechanism as deduced from such idealized models is sketched in Fig. 8 and described in the caption. Such single patterns were also determined in global ocean models, where the time scale is multi-millennial (Weijer and Dijkstra, 2003).

As a second example, consider the centennial variability which was found in a 1,500 years long simulation with the Kiel Climate Model (KCM) using present-day constant forcing conditions (Latif et al 2013). The observable used is the Southern Ocean Centennial Variability (SOCV) index, defined as the zonally and meridionally (from 50S-70S) averaged SST anomaly. The SOCV shows centennial variability for which the red noise null-hypothesis can be rejected. Analysis shows that convection in the Weddell Sea is crucial in causing the variability, with responses on sea ice extent and AMOC in turn affecting the convection.

In this case, it is more difficult to attribute a pattern to the variability,

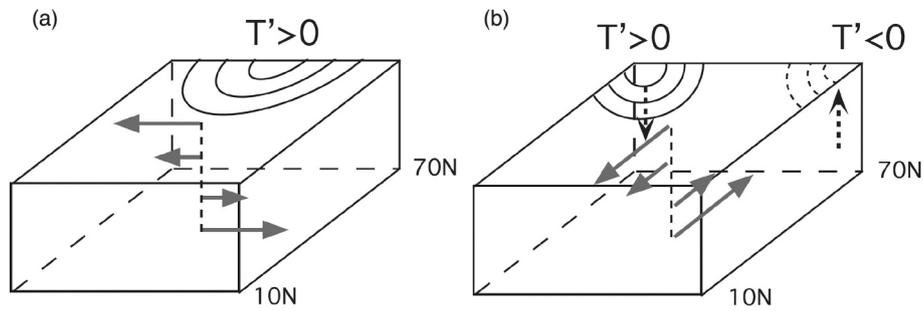

**Fig. 6.** Sketch to describe the mechanism of the AMO. A positive temperature anomaly causes changes in the zonal ocean circulation due to thermal wind balance. This causes the anomaly to propagate westwards causing a response in the meridional overturning circulation. The time scale of the variability is determined by the zonal basin travel time, details in Frankcombe et al. (2010).

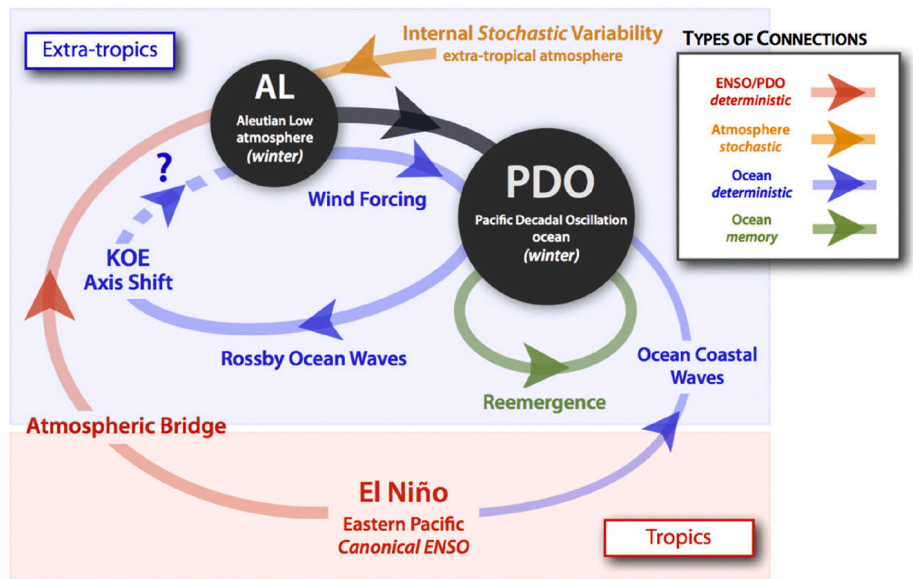

**Fig. 7.** Sketch of the mechanism of the PDO. SST anomalies can arise through reemergence of ENSO induced variability. Atmospheric noise excited Rossby waves which interact with the Kuroshio current, possible leading to path shifts leading also to SST anomalies, details in Newman et al. (2016).

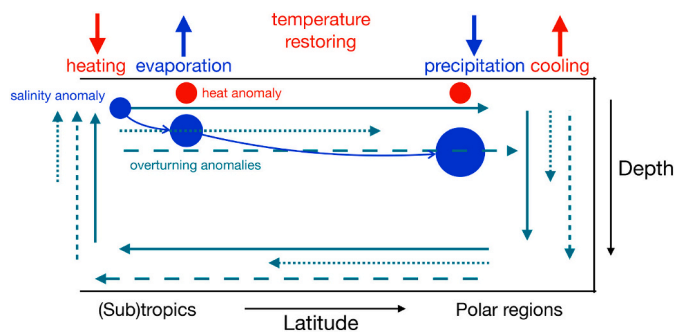

**Fig. 8.** Sketch to describe the mechanism of the Loop Oscillation. A positive salinity anomaly is propagating with the AMOC. While it is the evaporating region, it weakens the AMOC, remains longer in this region and hence is amplified. Next, in the precipitating region, it strengthens the AMOC, is shorter in this region and is amplified. Moreover, because of different damping of temperature and salinity anomalies, temperature-induced density anomalies appear, which are out of phase with those caused by salinity and hence cause the oscillatory nature of the variability. The timescale is determined by the propagation time of the salinity anomaly over the loop defined by the AMOC (details in Sevellec et al. (2006).

because no single pattern in an idealized model has been found causing this type of variability. However, there are idealized models showing the variability caused by transitions between convective and non-convective states (Welander, 1982). These changes can therefore best be described by 'Convection-Restratification' variability; when the restratification takes place through diffusive processes the variability has been called a deep-decoupling oscillation or a 'flush', (Winton and Sarachik, 1993). A sketch of the mechanism of the 'Convection-Restratification' variability is given in Fig. 9 (with a description in the caption). Here the time scale is dependent on the processes restoring the stratification. When this process is vertical mixing the time scale is millennial (Colin de Verdière, 2007) but when faster advective processes are involved, the time scale can decrease to centennia or even (multi)-decadal (Bars et al., 2016).

### 2.5.4. Dansgaard-Oeschger and Heinrich events

Very abrupt climate shifts between a cold glacial (stadial) state and a warmer (interstadial) state were first observed in Greenland ice core records (Dansgaard et al., 1993) and named Dansgaard-Oeschger (DO) events. The duration of the abrupt jumps is of the order decades, while the duration of the stadial - and interstadial periods is of the order a few centuries to a several millennia. When first discovered in ice core records their cause was unknown, but the apparent regularity of jumps at the millennial scale suggested a periodicity around 1500 years (Schulz, 2002). The regularity of the events has been interpreted in favor of forcing external to the Earth system, possibly solar origin (Rahmstorf, 2003), which could be amplified by AMOC and Arctic sea ice feedbacks Bond et al. (2001), suggesting interaction between external forcing and internal variability. On the other hand, it was shown that the distribution of waiting times between jumps is consistent with a random memoryless process (Ditlevsen et al., 2005), which would suggest no periodicity in the series of events (Peavoy and Franzke, 2010).

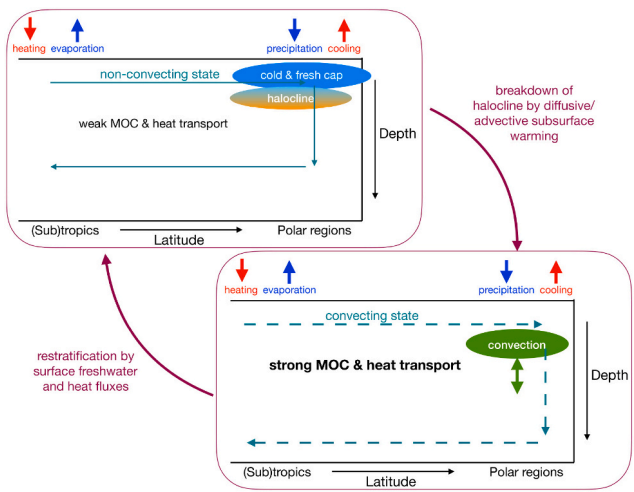

**Fig. 9.** Sketch to describe the mechanism of the Convection-Restratification variability. Starting in a non- or weakly convecting state, advective (wind-driven) or diffusive warming of the subsurface ocean causes convection and breaks down the halocline. The convective state increases the strength of the AMOC. Through advective and/or diffusive heat and salt fluxes, restratification occurs, which in turn reduces the density at the surface waters eventually causing the transition back to the non-convective state (details in Winton and Sarachik (1993).

The DO events are observed not only in the Greenland ice cores but in a suite of paleoclimatic records such as ocean sediment cores (Shackleton et al., 2000) and speleothems (Wang et al., 2001) spread across the Earth. Likewise, the DO events have counterparts in Antarctic ice cores (EPICA Community Members, 2006), connected via the empirical seesaw model (Stocker and Johnsen, 2003). Thus DO events seem to have an almost global extent.

The dynamics underlying the DO events are not known and several components of the climate system around the North Atlantic have been proposed to be involved: Variation of intermediate-size ice sheets (Zhang et al., 2014), ice shelfs Petersen et al. (2013) or change in sea ice as well as changes in the AMOC (Bond et al., 1997; Ganopolski and Rahmstorf, 2001; Vettoretti and Peltier, 2016).

Most of the proposed models for explaining the DO events involve two stable states, either as two branches of a slow manifold in a fast-slow self-sustained or forced oscillator system, as coherence or stochastic resonance or as a bimodal system with noise induced transitions. Either of the possibilities will reflect itself in the climate spectrum: In the case of an oscillating system the spectrum will show spectral peaks or resonances, while in the stochastic dynamics case, the spectrum is expected to be more continuous. From the records, the latter seems to be the case (Ditlevsen and Johnsen, 2010).

Some of the Greenland Stadials are sometimes qualified as Heinrich Stadials, because they include *Heinrich events (HEs)*, though this wording calls for some caution. In principle HEs should be defined on their own, independently of the Greenland stratigraphy. Within deep-sea sediments, Heinrich *layers* are characterised by exceptional abundance of ice-rafted debris (mainly quartz) in the so-called band of Ruddiman, a large zone covering the latitudes of $40^o$ to $50^o$N in the North Atlantic (Heinrich, 1988). The provenance of these debris has been studied in detail (Grousset et al., 2001; Hemming (2004)) and much work has been made also to use carbon and oxygen isotopes and other geochemical tracers to characterise the water hydrography before, during and after Heinrich events (Vidal et al., 1997, but see Crocker et al. (2016) for a more recent overview).

The debris comes from icebergs delivered by the surrounding ice sheets (North America and Fennoscandia), which are undergoing some form of catastrophic dynamics. These dynamics may be related to phenomena such as basal sliding or the rupture of ice shelves, but these phenomena are certainly part of a more complicated causal chain with self-amplifying loops. Among others the sea-level rise caused by iceberg calving has a potentially destabilising action on all ice sheets, including Antarctica. One of the immediate correlates of Heinrich events is a large and widespread reduction (but probably not full suppression) of the ventilation of North Atlantic intermediate waters, with various atmospheric consequences such as a southward shift of the intertropical convergence zone.

Heinrich Events have first been modelled as a stand alone self-sustained glaciological oscillation (MacAyeal, 1993), with a typical return time of about 7 kyr, roughly consistent with observations. As ocean and ice sheets have to be considered as part of the coupled system, and Dansgaard and Heinrich events may entertain some causal connection (Bond et al., 1997; van Kreveld et al., 2000). For example, the Heinrich Events may also be plausibly modelled as a non-linear resonance to Dansgaard-Oeschger oscillations (Alvarez-Solas et al., 2010). As for Dansgaard-Oeschger variability, there is enough indication that HEs occurred during previous glaciations, surely since Middle-Pleistocene Transition (Obrochta et al., 2014) but the stratigraphy is less well established.

Several long (millennial) GCM simulations have been performed under glacial climate conditions to study the Dansgaard-Oeschger oscillations. Vettoretti and Peltier (2016) used the CESM1 model and studied its variability under glacial forcing conditions after a perturbation associated with a Heinrich event was introduced. They found that very regular, millennial time scale oscillations appear of which the period slightly changes in time. Analysis of the model results leads them to suggest that a salt - oscillator (involving the AMOC and the sea ice distribution) is responsible for the variability. From the model results it is not clear whether the oscillations extend above a red noise background.

### 2.6. Interaction of internal variability and external forcing

A natural question to ask is how the internal variability and external forcing of the climate are interacting. Although for the diurnal and annual forcing this relation is clear, over long timescales this is a difficult question to answer. We focus our discussion to glacial cycles and longer timescale cycles where there has been some progress since Mitchell (1976).

#### 2.6.1. Glacial cycles

Through the natural history of the Earth at least five geological periods, including our present time, have been been characterized by a relatively cold climates and extended glaciations. The first well documented ice age is the Huronian at 2.4 Ga with several individual glaciations, of which one might have covered the world in an Snowball Earth event, possibly caused by depletion of methane from the atmosphere after oxygen levels rose at the Great Oxidation Event (Young, 2013; Kopp et al., 2005). The next and most dramatic is the Cryogenian 720-630 Ma in which two Snowball Earth events occurred (the 'Sturtian' and 'Marinoan' glaciations), with ice sheets reaching the equator. Two other ice ages, the Hirnantian (about 445 Ma) and late Paleozoic (360-260 Ma) happened prior to our present late Quaternary icehouse beginning with the emergence of the Antarctic ice sheet and finally the past 3 Ma Pleistocene glacial-interglacial cycles characterized by waxing and waning of Northern hemisphere ice sheets. Conditions for developing large ice sheets change on multi-millennial geological time scales.

In the Pleistocene the distribution of continents, the South Pole being land bound, the Arctic Ocean relatively isolated and the atmospheric greenhouse gas concentrations being low, made the climate susceptible to the changes in insolation due to the orbital forcing.

The main proxy record for the Pleistocene climate is the oxygen isotope ratio in benthic foraminifera in the deep sea sediments. This is a combined proxy for the deep sea temperature and the global ice volume. The latter since the isotopic fractionation favors light isotope oxygen and hydrogen in forming ice sheets, thus leaving the ocean with a higher concentration of heavy isotopes. The former comes from the temperature dependent fractionation in the biological formation of the foraminiferal shells. The isotope records from deep sea cores all over the world oceans are similar, and in order to eliminate local sources of noise and variations, Lisiecki and Raymo (2005) (LR05) constructed an average -or stack- of 57 records spread over the world oceans.

The influence of the changing insolation fields on the growth and decline of the glacial ice sheets can be represented by the annually averaged insolation at 65°N, approximately corresponding to the southern rim of the ice sheets, dominating the melting of the ice sheet. The past 3 Ma of this insolation curve and the LR05 record are shown in the upper panel of Fig. 10. The lower panel shows the spectral power within a 400 kyr running window (indicated by the bar). There is a strong weight around the 41 kyr band, similar to the period in the insolation curve. At the Middle Pleistocene Transition (MPT, gray band at 1200 - 800 ka BP) strong spectral weight around 100 kyr period builds up and has dominated glacial cycles until present time.

Beside the change in glacial cycle duration, the paleoclimatic record also shows a slow cooling trend through the past millions of years. The most plausible cause of this cooling is a decrease in the atmospheric $CO_2$ concentration, which in turn could make way for larger Northern Hemisphere ice sheets which would survive several insolation cycles (Lunt et al., 2008; Verbitsky et al., 2018). As mentioned before, the climate response to the astronomical forcing should not be expected to be linear as becomes obvious from the variable length of glacial cycles during the last few Ma. Several models, from more conceptual to ones based on more-or-less realistic description of coupling between dominant physical components have been proposed to account for the glacial dynamics (Saltzman, 1990).

The characteristic temporal asymmetry in the record (the saw-tooth shape) points to an interplay between fast and slow time scale dynamics. A generic model for such a behavior is a fast-slow oscillator, as proposed in several models (Crucifix, 2012). A periodic forcing in such systems can lead to synchronization i.e. frequency locking between the internal oscillator frequency and the external forcing frequency (Gildor and Tziperman, 2001). The MPT would then be a manifestation of a change from a 1:1 over a 2:1 to a 3:1 frequency locking perhaps triggered by a slow change in an external parameter, such as the $CO_2$ concentration, (Clark et al., 2006; Nyman and Ditlevsen, 2019). An alternative interpretation is that the MPT reflects a change in the structure of the slow manifold in the fast-slow system (see Section 4) in such a way that a new branch, the deep glacial state, becomes accessible, (Paillard, 1998). A generic way of describing this is through a transcritical bifurcation (Ashwin and Ditlevsen, 2015). Both of these proposed mechanisms seem to give a better match to the record than the original proposal by Saltzmann of a Hopf bifucation occurring in the system at the MPT Saltzman and Maasch (1990). It has also been suggested that the MPT could occur without the need of a slowly varying parameter (or external forcing), where ice-sheet dynamics can (spontaneously) exhibit period-doubling (Raymo, 1997; Verbitsky et al., 2018; Verbitsky and Crucifix, 2020) or because of a combination of delayed feedback and bistability (Quinn et al., 2018).

The narrow peak spectrum near the astronomical periods proposed by Mitchell is in principle a time independent average of the spectral weights. A more informative picture of the climate variability on glacial time scales, reflecting the climate susceptibility and response to astronomical forcing would be represented by a wavelet expansion or a finite window periodogram as shown in Fig. 10.

### 2.6.2. Long timescale variability: biogeochemical and supercontinent cycles

Although the records become sparser as the timescales become longer, various studies have presented evidence for interaction between intrinsic and forced long timescale climate variability. Internal (unforced) periodic variability can arise from the interplay of positive and negative feedback mechanisms within the Earth system. Potential sources of low frequency periodic external forcing include orbital forcing, mantle convection and plate tectonic cycles, and possibly galactic cycles (e.g. rotation around the galactic centre and passing through arms of the galactic spiral).

A candidate for intrinsic very low-frequency variability are the succession of Mesozoic ocean anoxic events (OAE) discussed in Handoh and Lenton (2003). There is evidence that peaks of phosphorus (P) accumulation are present in marine sediments exhibiting an approximate 5-

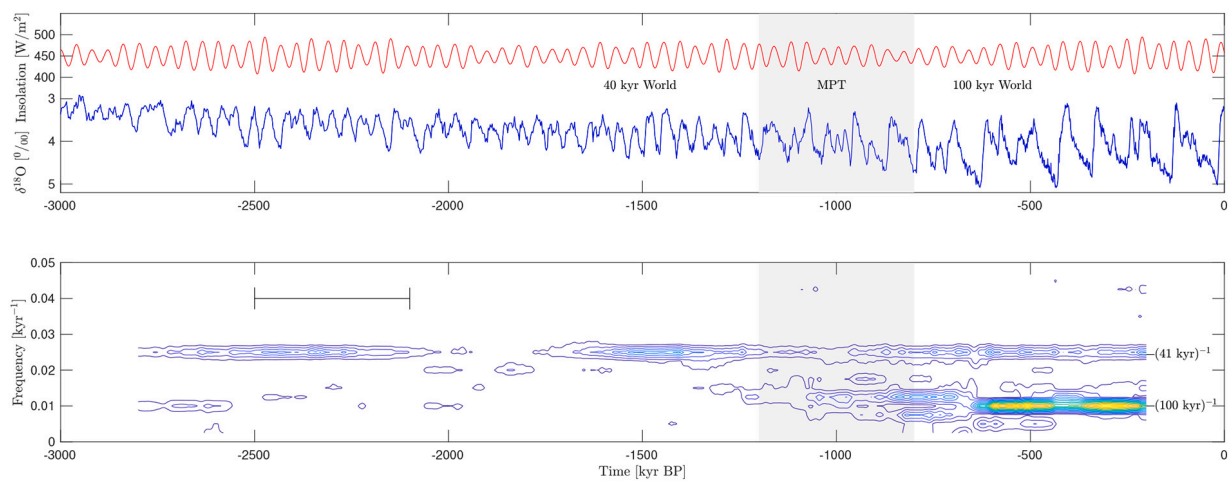

**Fig. 10.** Top panel: The LR05 stacked benthic foraminifera isotope record of the Pleistocene (blue). The record is a proxy for global ice volume. The red curve is the annually averaged insolation at 65N latitude, which is governed by the obliquity, i.e. the tilt of Earth's axis with respect to the plane of the orbit. Prior to the Middle Pleistocene Transition (MPT) the glacial cycles had an approximately 40 kyr period, thus called the "40 kyr World", very similar to the 41 kyr obliquity period. At the MPT (gray band) the duration of the glacial cycles increases into the 100 kyr World of the late Pleistocene. The bottom panel shows the spectral power of the record within a running 400 kyr window (indicated by the black bar). The record has a pronounced spectral peak at $(41 \text{ kyr})^{-1}$ through the whole record, while a broad peak around $(100 \text{ kyr})^{-1}$ emerges after the MPT.

6Myr periodicity. In Handoh and Lenton (2003) this is associated with self-sustaining oscillations in a model of coupled oceanic $N$, $P$, $C$ and $O_2$ biogeochemical cycles. The potential for oscillation arises because of the interaction between a fast positive feedback in which ocean euxinia (Meyer and Kump, 2008) enhances the recycling of phosphorus fuelling more euxinia, and a slower negative feedback whereby increasing atmospheric $O_2$ tends to re-oxygenate the ocean. A secular increase in phosphorus weathering rate can tip the system (through a Hopf bifurcation) into slow, self-sustaining oscillations. Watson et al. (2017) simplifies this model to explore the possibility that the anthropogenic increase in $P$ weathering due to climate change and direct $P$ mining is a major perturbation that could trigger an ocean anoxic event. These and similar mechanisms have been proposed for the Neoproterozoic-early Paleozoic (Mills et al., 2011; Alcott et al., 2019). Furthermore, Wallmann et al. (2019) propose a spatial redox see-saw oscillation on shorter timescales within individual Mid-Cretaceous oceanic anoxic events.

Recently there has been a burgeoning of studies considering the slowest components of orbital forcing, which are increasingly being used for sequence stratigraphy, and are showing up in deep time records of climate ($\delta^{18}O$) and the carbon cycle ($\delta^{13}C$). The eccentricity of the Earth's orbit is modulated at low frequencies that extend beyond 405 kyr Berger (1978a) to include 2.4 Myr (Kocken et al., 2019), 4.86 Myr (Matthews and Al-Husseini, 2010), and 9 Myr (Boulila et al., 2012; Martinez and Dera, 2015), see Fig. 4a. Obliquity is also modulated at low frequencies, e.g. 1.2 Myr. Various models suggest (linear or nonlinear) amplification of these orbital frequencies in the carbon cycle to explain their pronounced appearance in the $\delta^{13}C$ record (Paillard, 2017; Pälike et al., 2006). The long time-scale response to eccentricity can also show up in glacial dynamics as has been suggested around the Eocene-Oligocene transition (DeConto and Pollard, 2003). A possible candidate for low-frequency orbital pacing of an internal oscillator in long-term climate dynamics and the carbon cycle, is the build-up and draining of a methane hydrate/clathrate 'capacitor', particularly during the Paleogene (Lunt et al., 2011).

Other frequencies that are not obviously related to Earth's orbit also show up in long records, e.g. Cenozoic $\delta^{13}C$ has 27 Myr variability (Boulila et al., 2012), and a number of proxies over Phanerozoic time, including ones for atmospheric $CO_2$, as well as the fossil record exhibit 26-30 Myr periodicity of uncertain origin (Prokoph et al., 2000). This has recently been linked to possible periodicity in sea-floor spreading and ocean crust carbon cycling (Müller and Dutkiewicz, 2018). A 60 Myr periodicity is well known in fossil biodiversity, and has also been detected in the strontium isotope record, and large-scale sedimentation (Melott et al., 2012). This might be linked to the emplacement of large igneous provinces (LIPs) associated with mantle plumes, which is argued to be periodic at 64.5 Myr (Prokoph et al., 2013).

One of the longest timescales postulated for internal variability is that of the tectonically-driven supercontinent cycle (Young, 2013; Condie, 2016) of repeated aggregation and dispersion of the continents: the most recent being Pangea (450-320Myr), then Rodinia (1000-850Myr) with supercontinents before these on roughly 500Myr intervals. This approximate timescale of what is called the "Wilson cycle" is set by the tectonic spreading rate and the circumference of the Earth. Both the accumulation and dispersal phases of the supercontinent cycle are associated with enhanced volcanic/metamorphic cycling of carbon and hence with 250 Myr periodicity in the long-term $CO_2$ record. Such changes will clearly have major influence on all aspects of the physical processes driving the climate, but given the uncertainty of records over these timescales much of this is likely to remain speculative.

## 3. Novel data analysis techniques

Subsequent to Mitchell (1976), wavelets (see e.g. Fig. 4) have been developed and established themselves as a commonly used data analysis technique to extract frequency-domain information from time series that may have non-stationary variability. Stationarity of time series in paleoclimate data records is generally the exception rather than the rule: they often have a substantial secular trend and the oscillatory component usually contains variations acting on a wide range of time scales. Furthermore, the oscillatory component often exhibits nonlinearity and nonstationarity. On the other hand, many classic time series analysis techniques (e.g. Fourier) assume linearity or stationarity of the data. These issues create challenges for the analysis of paleoclimate records and for drawing inferences about the underlying processes based on those analyses. In this section, we highlight wavelet and more recent sparse-decomposition techniques that aim to overcome some of these problems.

In a classic spectral analysis of a time series, the record is split into a secular trend and an oscillating residual. The trend is determined by an a priori choice of model - typically either a regression to some order polynomial or a convolution with some choice of smoothing window. The choice of model for the trend can have some impact on the subsequent analysis of the oscillatory residual. The residual is then represented by a decomposition over a set of basis functions.

$$d(t) = y(t) + x(t) = y(t) + \sum c_n \phi_n(t) \qquad (6)$$

where $d(t)$ is the original time series, $y(t)$ is a secular trend, $x(t)$ is the oscillatory residual, $\{\phi_n(t)\}$ is a set of basis functions and the $c_n$'s are the coefficients used to express $x$ in terms of the set of the $\phi_n$'s. It is useful to view any decomposition as a combination of two components: a *dictionary* and an *algorithm*. The dictionary is the full collection of potential basis functions, chosen a priori. The algorithm, either using the entire dictionary or selecting a proper subset of the dictionary, determines the unique coefficients used to express the given record in terms of the chosen subset of the dictionary.

For a classic Fourier analysis, the dictionary consists of a set of sinusoidal functions where the frequencies are chosen to create an orthogonal basis over the space of possible time series of the same length as the record being analyzed. However, from a statistical perspective, a data record should be viewed as a sample of an unknown population both in terms of the finite length of the record and finite time step size (Press et al., 2007; Wilks, 2006). There exists a wealth of techniques designed to address this fundamental statistical question – the goal being to provide a statistically robust estimate of the true (unknowable) power spectrum. These techniques include the Maximum Entropy Method, Multi-taper Method, Singular Spectrum Analysis and myriad others (Ghil et al., 2002; von Storch and Zwiers, 1999). In general, they can be viewed as finding a Fourier decomposition of a model for the data record where the model can be chosen a priori or data-adaptively.

The strengths of the Fourier approach include a rigorous mathematical foundation provided by harmonic analysis, the existence of a unique representation in terms of an orthogonal basis, a deep understanding of the statistical nature of the analysis and decades of experience within the scientific community in interpreting the analyses in order to make inferences about the processes which created the time series. The weaknesses of the Fourier approach include the assumptions of linearity and stationarity and the a priori choice of basis functions, which cannot change in time. The latter property means that a so-called 'global' projection onto the basis is used and consequently results in a 'global' frequency analysis. While any time series, stationary or non-stationary, can be represented uniquely, when the record is statistically non-stationary, the dominant time scales found in the 'global' analysis may not provide a good match with the time scales of the underlying processes. Even if the record is statistically stationary, when the shape of the regular oscillations is not close to sinusoidal, the number of orthogonal sinusoidal modes needed to explain a non-sinusoidal signal is large. Finally, the construction of the decomposition via projection onto an orthogonal basis is a fundamentally linear perspective. As such it may not provide a natural or efficient representation of a signal containing a high degree of nonlinearity.

### 3.1. Handling non-stationary data

There are a number of techniques designed to address the issue of non-stationarity; their goal is to provide a time-frequency analysis. Two of the most prevalent families of such techniques are windowed Fourier analysis and wavelets. In windowed Fourier analysis, a Fourier analysis is applied to a sliding window whose length is substantially shorter than the given record; a global projection is applied to each windowed subset of the data (Mallet, 1998). Effectively, there is a trade-off between the temporal resolution and the low-frequency spectral resolution. The issues involved in projection onto an a priori, orthogonal basis then remain. Huybers (2007) applies such a windowed Fourier analysis to a Pleistocene record of glacial cycles.

In a classic wavelet analysis, the dictionary is a set of orthogonal functions with compact support or approximately compact support – identically zero or vanishing small outside of a finite interval (Daubechies, 1992; Mallet, 1998). The interval of support can be rescaled to capture different time scales and the center of interval can be moved to address non-stationarity. The algorithm determining the decomposition coefficients is still a global projection, but the compact support implies that the resulting coefficients depend on the width and location of interval of support, i.e., the power of an oscillation is represented as a function of the time scale of oscillation (or frequency) and of a specific temporal window. A dictionary of orthonormal functions can be constructed from a "mother" wavelet, $\Psi(t)$, via dilation and translation:

$$D_W = \left\{ \Psi_{a,b}(t) \right\} \quad \text{where} \quad \Psi_{a,b} = \frac{1}{\sqrt{a}} \Psi\left(\frac{t-b}{2a}\right) \tag{7}$$

where the $a$'s and $b$'s are chosen to create an orthonormal basis on the space of time series of the given length. There are a large number of "mother" wavelets shapes that are typically used to construct the dictionary, such as the Meyer or Mexican-Hat wavelets. The algorithm uses the entire basis, i.e.

$$x(t) = \sum c_{a,b} \Psi_{a,b}(t) \quad \text{where} \quad c_{a,b} = <x(t), \Psi_{a,b}(t)> \tag{8}$$

where $c_{a,b}^2$ represents the power of oscillations on a time scale determined by $a$ for a window whose center is $t = b$ and whose width of compact support is determined by $a$. Fig. 10 provides an example of using a wavelet analysis to analyze the glacial cycle record of Lisiecki and Raymo (2005); Lisiecki (2010) for the Pliocene and Pleistocene.

Classic wavelet analyses have some of the same strengths as Fourier techniques such as a deep underlying mathematical theory, and a unique representation of any record over a priori choice of wavelet family. In addition, since the projections are effectively local, they add the key ability to represent non-stationary data more clearly. However, they still retain the weaknesses inherent in using projections onto an a priori choice of orthogonal functions. It should be noted that there is a rich collection of specialized wavelet techniques to address various aspects of time series analysis. For example, there are data-adaptive wavelet techniques which use a large dictionary consisting of multiple wavelets families. Such an algorithm first selects an optimal set of wavelets then determines the required decomposition coefficients for that subset of the dictionary (Mallet, 1998). This idea of starting with a very large dictionary, selecting a small optimal subset in a data-adaptive manner and then determining a decomposition in terms of this small subset is at the core of sparse decomposition techniques.

### 3.2. Sparse decomposition techniques

In a classic Fourier or wavelet analysis, the number of modes needed is large and determined a priori. Since the entire orthogonal basis is used in the decomposition, the number of modes is the same as the length of the data record. For example, a 2 Myr record, sampled at 1 kyr intervals, has 2000 data points; so a Fourier analysis requires a basis consisting of 2000 sinusoidal functions – 2 modes for each of 1000 frequencies

(modulo a slowly varying component). Generally, unless there is a close match between the elements of the dictionary and the characteristic of the observed oscillations, a large number of the modes must be retained to capture a large fraction of variance of the time series.

In a sparse decomposition, the goal is to find a decomposition of a given data time series over a relatively small number of modes. If we substantially increase the number of elements in the dictionary, we can design data-adaptive algorithms to determine a low-dimensional subspace containing the time series which is spanned by a small number of dictionary elements (not necessarily mutually orthogonal elements.) To do so, we must relax the requirement that the dictionary be a basis or be an orthogonal set. For example, we can construct a large dictionary using *intrinsic mode functions* (IMFs) – functions which locally oscillate about a zero mean with a slowly varying amplitude and frequency:

$$D_I = \left\{ \phi_n(t) \right\} = \left\{ a_n(t) e^{i\theta_n(t)} \right\} \tag{9}$$

where, for each mode, the amplitude, $a_n(t)$, and the instantaneous frequency $f_n(t) = \dot{\theta}_n(t)$ vary slowly compared to the timescale of oscillation, approximately $2\pi/f_n(t)$. Note that an entire Fourier dictionary, $D_F$, is a special case of $D_I$ in which $a_n$ is constant and $\dot{\theta}_n(t) = f_n t$ for $f_n = nf_0$. The algorithm can then select an optimal subset of the dictionary with which to construct the decomposition, but the criteria for determining the optimal subset can vary between algorithms. So even for the same dictionary, the decomposition can be quite algorithm dependent.

The key strength of sparse techniques is the efficiency of the decomposition – the reduction to a relatively small number of modes can simplify the task of drawing inferences from the data as to the potential underlying processes. Over the last couple of decades, there has been a growing interest in the development of both sparse decomposition techniques and an underlying mathematical theory (Hou and Shi, 2011). There is a wide variety of techniques such as those based on data-adaptive wavelets or matching pursuits. The variety of algorithms allows different analyses to be chosen to address various issues inherent in analyzing any particular data record; eg., extraction of instantaneous frequencies from a complicated signal, poor scale separation, nonlinearity, nonstationarity, and/or poor signal to noise ratio (Daubechies et al., 2011; Mallet and Zhang, 1993; Hou and Shi, 2016).

### 3.3. Ensemble empirical mode decomposition

Empirical Mode Decomposition (EMD) (Huang et al., 1998; Huang and Wu, 2008) provides a sparse decomposition over $D_I$, the dictionary of intrinsic mode functions defined in Equation 9, plus a final residual secular trend mode. The algorithm is a sifting process consisting of a nested pair of iterations and is designed to decompose the data into a sequence of IMFs which capture the variability on different time scales sorted from fastest to slowest. The first IMF is constructed by isolating the fastest oscillations about a locally defined mean by iteratively removing the average of the upper and lower envelopes of the original record. The first IMF is then subtracted from the original data, effectively a smoothing operation, and the process is repeated to construct the second IMF. The removal of the second IMF is also a smoothing operation, just on a longer time scale than the removal of the first IMF. This outer iteration is subsequently repeated until the remaining smoothed data contains no further oscillations; this remaining non-oscillatory mode is called the residual. One benefit of EMD is that there is no need to detrend the data prior to analysis. A data-adaptive secular trend is part of outcome.

Ensemble Empirical Mode Decomposition (EEMD) is a noise-assisted refinement of EMD, which substantially improves the coherence in instantaneous frequency for each IMF, thereby facilitating the physical interpretations of individual IMFs. EEMD is the average of the EMD analyses for an ensemble of time series, each of which consists of the original time series plus a distinct random white-noise series. (Wu and Huang, 2009). Both EMD and EEMD can be viewed as a collection of bandpass filters – a filter bank (Flandrin et al., 2004). Furthermore, for a pure white noise signal, there is an inherent period doubling between sequential modes; so, for a signal of length $N$, there is an expectation that the decomposition will have approximately $\log_2 N$ IMF's plus the non-oscillatory residual – a large reduction from the $N$ modes needed in a classic Fourier analysis. Finally, since the IMFs are elements of $D_t$, a class of mono-component oscillations, a Hilbert transform can be applied to extract the instantaneous frequency for each mode and also a spectral power estimate as a function of that instantaneous frequency and time, i. e., a time-frequency analysis.

EEMD is a data-adaptive, local analysis in which no projections are used to construct the IMFs. As such, it is well suited to find a sparse decomposition for non-stationary and nonlinear time series. It is particularly well suited for isolating a fast oscillation modulated by a slowly varying amplitude. It is also capable of cleanly extracting a subdominant fast oscillation in the presence of a dominant slower oscillation if there's sufficient scale separation between them. In addition, there is no need to use an a priori model to detrend the data prior to analysis; EEMD provides a data-adaptive secular trend.

### 3.3.1. An EEMD analysis of a glacial cycle record

As an example of how EEMD can extract changing periodicity and transitions in a non-stationary, nonlinear setting we give an analysis of the last 2.5 Myr benthic $d^{18}O$ of Lisiecki and Raymo (2005), hereafter denoted as the "LR05" stack. Huybers and Wunsch (2004) constructed a similar stack for the late Pleistocene using an depth-derived age model, mapping depth to time, based on models of sedimentation rates and therefore largely devoid of assumptions of pacing to orbital tuning in the tuning of the age model. Huybers (2007) extended this astronomically untuned stack into the early Pleistocene and analyzed it using a windowed Fourier analysis in which there appeared to be a transition of power from the 40 kyr time scale to the 100 kyr time scale. Based on this analysis and the observation that the timing of large warming events is paced by obliquity, Huybers (2007) proposed a "skipped-obliquity" model to explain the observed transition in the Fourier analysis, On the other hand, Lisiecki (2010) performed a cross-spectral wavelet analysis of eccentricity and a modified LR05 stack with a new depth-derived age model and concluded that the 100 kyr cycles were paced, but not necessarily forced, by the variations in eccentricity. The choice of different analysis techniques emphasized different features of the data records in comparison to the astronomical forcing. EEMD can provide a distinct, complementary viewpoint.

Fig. 11 shows an EEMD analysis of the modified LR05 stack from Lisiecki and Raymo (2005) during the Pleistocene. The original data record is shown along with resulting 10 IMFs and the non-oscillatory residual. A data-adaptive trend, consists of the last 3 IMFs plus the residual is shown superimposed on the original data. The choice choice of IMFs used in the trend was made a posteriori to capture the variability on time scales longer than 400 kyr.

The first 2 IMFs capture the variations on the fastest time scales, but the amplitude is negligible. IMF 3 captures variability on the precessional time scale, ~20 kyr. IMF 4 captures variability on the obliquity time scale, ~40 kyr, and IMF captures variability on the ~100 kyr time scale. The key features, captured in IMFs 4 and 5, are readily apparent:

- IMF 4 shows that in the early Pleistocene (before 1.25 Mya), the record is dominated by variations on the obliquity time scale – precessional time scale oscillations, captured in IMF 3, are relatively small. We can also observe the slow modulation of the amplitude of the 40 kyr cycles captured within this single mode.
- IMF 5 captures the most striking feature - the appearance at approximately 1.25 Mya of a new slow oscillation which grows to dominate the 40 kyr oscillations during the last 600 – 700 kyr. However, during the late Pleistocene, the 40 kyr oscillations clearly persist albeit with less regularity than during the early Pleistocene when 40 kyr is the dominant time-scale of variability.

The variations captured in IMFs 7 and 8 during the late Pleistocene occur on ~200 kyr and ~400 kyr time scales; they are smaller amplitude that the 100 kyr cycles and appear to be paced (with skipping) by the 100 kyr cycles. These may not be independent signals, but simply mathematical artifacts of the inherent doubling nature of EEMD, analogous to the higher harmonics that occur in the Fourier analysis of a non-sinusoidal signal. Subsequent analyses of the slow variability can be done using IMF 5 or a sum of modes 5 or 6, 5, 6, and 7; such a summation of modes represents an a posteriori choice of a broader

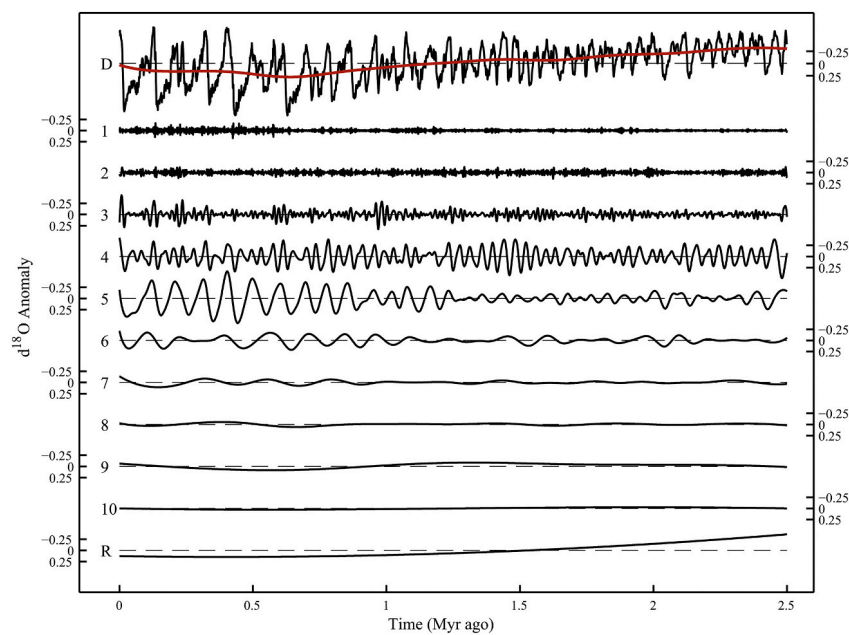

**Fig. 11.** EEMD analysis of a $d^{18}O$ record of glacial cycles during the last 2.5 Myr. D denotes the original data record from modified LR05 stack. 1 – 10 denote IMFs 1 – 10. R denotes the residual. The red curve in D is a data-adaptive trend consisting of the sum of IMFs 8 – 10 and the residual. The analysis was performed using 2500 ensemble members and a white noise distribution with standard deviation equal to 50% of the standard deviation of the original data.

bandpass filter for subsequent analyses.

This example demonstrates that EEMD can provide a complementary perspective to the spectral picture in Fig. 10 of a data record. The algorithm's sifting process, with its inherent period doubling, is well suited for decomposing signals consisting of variability on multiple time scale if those scales are sufficiently separated. The order of the outer iteration in the algorithm – isolating faster oscillations first – is well suited to extracting a sub-dominant fast signal in the presence of a dominant slow signal. And the algorithm's use of local analysis without resorting to projections is well-suited to analyzing signals which are non-stationary or nonlinear. The choice of a dictionary consisting of mono-component signals allows the extraction of a power vs time and instantaneous frequency spectrum via a Hilbert transform. Finally, individual modes or summations of modes, representing various bandpass filters of the data, can be analyzed further in the temporal domain.

## 4. Developments in the theory of forced systems

A spectral decomposition of a signal is most relevant when there is stationary forcing of a linear system, but there is plenty of evidence that both assumptions are only approximately relevant in the climate system and only over certain timescales There is now a recognition that chaotic behaviour, multistability (multiple attractors) and tipping points are a common feature of many systems with nonlinear feedbacks.

Trends can be filtered out using techniques such as detrended fluctuation analysis (Kantelhardt et al., 2002), but in general one needs to identify periods of stationarity before applying a spectral approach and there are dangers in a naive approach that assumes linear response (Gottwald et al., 2016). Change-point/structural break detection methods (Gallagher et al., 2013; Reeves et al., 2007) are often employed in econometrics and are now emerging as tools to deal with nonstationary climate data (Mudelsee, 2019)).

### 4.1. Synchronization, resonant response and pacing effects

It is natural to consider the extent to which external forcing be responsible for observed climate variability. At one level this seems to be quite a simple question of matching the spectra of the forcing to the spectra of the climate and noting the presence of peaks that occur in both (see for example Fig. 2). This approach works well in cases where:

(a) The forcing is well characterized (fortunately the astronomical forcing is accurately known over the past few million years),

(b) The climate response is well characterized (this is clear for some proxies, but essentially limited to local records at a small number of points that may give global proxies).

(c) The climate system can be well approximated as a forced linear system.

The linear approximation (c) will become poor if there are internal timescales, periodicities or nonlinearities in feedbacks that can influence the response in a number of ways. This can cause the robust appearance of a wide range of spectral phenomena in the absence of any forcing. The response can "synchronize" or more generally have "nonlinear resonance" with the forcing. This may result in a range of phenomena that complicate the linear approach above - including exact and partial frequency locking at multiples or fractions of the forcing periodicities. It is also known that nonlinear resonance can result in multi-stability and history dependent responses. It can also result in chaotic and effectively unpredictable response to non-chaotic forcing (Pikovsky et al., 2001), even for a system that shows no chaos in the absence of forcing, and even in the presence of simple periodic forcing; see for example the ENSO example Section 2.5.2, and the Pleistocene ice age cycle forced by orbital variability in Section 2.6.1.

More precisely, consider a system whose dynamics are goverened by an ODE that is forced by an external signal

$$\frac{d}{dt}x = f(x, \Lambda(t)) \tag{10}$$

where $f$ depends on the state $x \in \mathbb{R}^d$ and some "forcing" $\Lambda(t)$. For a stable linear system

$$\frac{d}{dt}x = Lx + \Lambda(t) \tag{11}$$

with $L$ a fixed stable linear operator, input-output theory of linear systems (e.g. Hinrichsen and Pritchard (2005)) gives a precise characterization of the the response $x(t)$, even in the case that $\Lambda(t)$ varies in a non-stationary manner. Indeed, one can write the solution

$$x(t) = x(0)e^{Lt} + \int_0^t e^{L(t-s)}\Lambda(s)\,ds \tag{12}$$

which means that the steady response is a convolution integral with no dependence on initial condition. This is a special case of the pullback attractor (Kloeden and Rasmussen, 2011) for a forced nonlinear system in the case where there is a single attracting trajectory.

For nonlinear climate models with general forcing, various studies (Chekroun et al., 2011; Ghil, 2015; Drótos et al., 2015; Ghil and Lucarini, 2019; Kaszás et al., 2019) have applied concepts from the theory of non-autonomous dynamical systems (Kloeden and Rasmussen, 2011) to understand aspects of behaviour. In this case the pullback attractor can consist of many trajectories that explore some time-dependent subset of phase space, also called a snapshot attractor Romeiras et al. (1990). For nonlinear systems of the form (10) with stationary forcing, this pullback attractor can have positive Lyapunov exponents and sensitive dependence on initial conditions even in the unforced case. Quasiperiodic forcing still has comparatively low complexity and little predictability, where $k$-frequency quasiperiodic forcing corresponds to forcing by a function of the form

$$\Lambda(t) = P(f_1 t, \ldots, f_k t) \tag{13}$$

such that $P$ is smooth and periodic with period $2\pi$ on all $k$ variables and the $f_i$ are the angular frequencies present. For typical choices of $f_i$ one can expect all frequencies are rationally independent of each other. Already with two independent frequencies, strange non-chaotic attractors may appear (Feudel et al., 2006) where the response does not have smooth dependence on phase differences.

Even for periodic forcing of a system where there is only a single stable periodic attractor, there can be a wide range of responses (Pikovsky et al., 2001) depending on relative frequency, amplitude of forcing and details of the nonlinearities involved. This includes periodic, quasiperiodic or chaotic response. For example, De Saedeleer et al. (2013) and Ditlevsen and Ashwin (2018) consider mode locking in response to periodic forcing in Pleistocene ice age models while (Ashwin et al., 2018) find that many such models can show chaotic response to periodic forcing.

For quasiperiodic forcing of Pleistocene ice age models, both strange non-chaotic (Mitsui and Crucifix, 2017) and chaotic responses (Ashwin et al., 2018) have been found. Note that many of the effects found in forced periodic systems can similarly appear as nonlinear resonances in systems that may not be periodic in the absence of forcing: indeed, in the presence of forcing there can be a cross-over between synchronization and resonance, as outlined in Marchionne et al. (2018).

### 4.2. Towards a phase space view of climate variability?

Notwithstanding the importance of the spectral view of variability, it does have limitations, especially in representing sudden shifts in climate associated with "tipping points". In such cases onlinear dynamical systems provide a powerful paradigm in which to view complex systems such as the climate and there are texts that specifically develop and apply these ideas in the climate system (Kaper and Engler, 2013; Dijkstra, 2013). Multiscale systems are mathematical models that resolve processes on many timescales. In many such systems (where dissipation is important) only the slowest processes are apparent for most of the time, but there can be short periods of time where the fast processes come into play. This approach is particularly useful in cases where there is a clear separation of timescales into fast and slow. For a stochastically perturbed equilibrium this corresponds to there being a spectral gap in the linearized system.

There are complementary approaches for fast-slow decomposition; if the fast process leads to stable equilibrium behaviour it is possible to adiabatically eliminate the fast processes. On the other hand, localized (unobserved small scale or, in model context, unresolved) fast processes that are chaotic lead to stochastically parameterized models (Berner et al., 2017). In general, these multiscale systems can be modelled using homogenization approaches such as Mori-Zwanzig (Mori et al., 1974; Zwanzig, 2001; Pavliotis and Stuart, 2008) applied to climate models (there is an extensive literature in this area, for example Majda et al. (2002, 2006); Gottwald et al. (2017); Falkena et al. (2019)).

If the fast system reaches an equilibrium then adiabatic elimination of the fast processes gives a description in terms only of the slow variables. More precisely, such a fast-slow system has a "slow manifold" on which the fast dynamics is "slaved" to the slow, other than remaining fluctuations on the fast timescale. However, in the presence of nonlinearities, this slow manifold may have folds, and passing over the fold can lead to a change point in the time series and a sudden readjustment of the fast processes: this provides a clear model where there can be both slow fluctuations, trends and fast transitions (critical transitions/tipping points) that lead from one part of the slow manifold to another. Fig. 12 illustrates this schematically: note that this can lead to stationary dynamical behaviour that can be well-modelled spectrally for periods of time, but where stationarity breaks appear as a new branch of the slow manifold is approached. Over larger timescales, stationarity may return: for example in simple nonlinear models of Pleistocene ice ages there may be regular "relaxation" oscillations between the parts of the slow manifold (Crucifix, 2012). The structure of these oscillations may gradually or abruptly change over longer timescales (Ashwin and Ditlevsen, 2015; Ditlevsen and Ashwin, 2018; Nyman and Ditlevsen, 2019).

A common way to move up the climate modelling hierarchy is to identify fast processes producing some degree of chaotic mixing and replace them by either an averaged or a stochastic perturbation of the slow processes. For example, a common way to model the Atlantic Meridional Overturning Circulation (Section 2.5.3) is to view the ocean/atmospheric turbulence as providing diffusive transport and/or a stochastic perturbations to a large scale flow (Dijkstra, 2013).

There are clearly limitations to either of these viewpoints: the required separation of timescales may not be consistent for all dynamics and may enter regions of phase space where timescales may cross over, for example in land ice dynamics where typical growth occurs very slowly, but ice sheet instabilities may progress on a much faster timescale.

## 5. Discussion and synthesis

This paper has revisited the climate spectrum of Mitchell (1976) which is reproduced in Fig. 1. Indeed, Fig. 2 gives an updated version that has been drawn based on the now available (more than at Mitchell's times) data and understanding of climate variability. While it is still not possible - and in fact may never be - to quantitatively estimate the spectrum from a single continuous time series, there are a number of notable additions in the updated version. First of all, peaks in the spectrum in Fig. 2 are drawn as deviations from a $f^{-1}$ background spectrum, and we indicate regions where a specific scaling of the background spectrum may apply according to the scaling regimes proposed (Lovejoy and Schertzer, 2013a). In the original MS (Fig. 1), the background spectrum follows more white noise characteristics (flat background) on the longer time scales. Externally forced peaks on very long time scales such as the one with 200-500 myr periodicity have much less spectral density in the updated spectrum, because the (forcing) periodicities are uncertain and it is impossible to test whether the response would show up above the background spectrum. Some other apparently forced spectral peaks have disappeared in the updated spectrum due to lack of evidence (e.g. the Brier forcing at 27 years). Moreover, in the time scale range between interannual and millennial the updated spectrum shows (i) some more broad peaks, e.g. related to ENSO, that were absent in the MS, and (ii) overall a higher spectral density related to internal modes of variability (some of which may not show up above the background spectrum for the global mean temperature). We highlight progress in estimating and interpreting the climate spectrum in a number of areas:

- A great improvement in quantity and quality of climate data both from observations covering about 150 years and from proxy records extending further back in time. Moreover, the tools needed to construct high resolution computational models of the earth system have undergone tremendous improvement.
- A better overview and understanding of the presence of interannual, multidecadal, centennial and millennial variability.
- A better understanding of Pleistocene climate oscillations, e.g. the glacial cycles, which have undergone a change in periodicity at the Mid Pleistocene transition.
- A recognition that various parts of the earth system may spontaneously oscillate, pass through tipping points or exhibit hysteresis due to nonlinear feedbacks.
- A better understanding of the prevalence of chaos and variability spontaneously appearing in nonlinear systems which suggests that apparently stochastic variability may in some cases be due to relatively low-dimensional processes.

Already at the time of Mitchell (1976), Hays et al. (1976) suggested that astronomical forcing is behind the observed oscillations in glacial cycles of the late Pleistocene in that it "paces" these cycles, more specifically by determining when deglaciations may occur. Although there is still no precise agreed interpretation of what pacing might be, a lot is known about synchronization between forcing and response in nonlinear systems, or synchronization between various internal components of a system, especially if those internal components have stable periodicity (see Section 4).

On the one hand, the emergence of high performance computing and vast improvement in availability and resolution of current and past climate data have revolutionized climate science since Mitchell (1976).

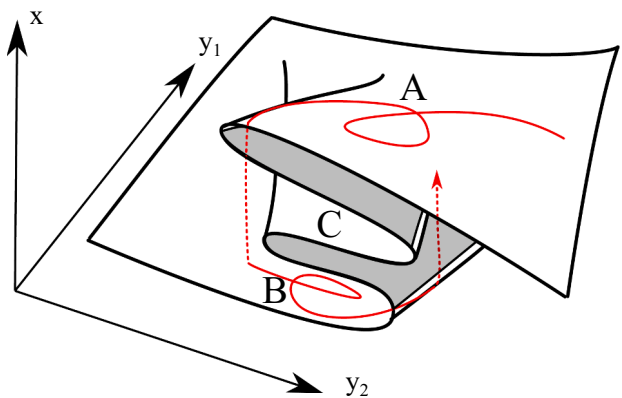

**Fig. 12.** Schematic diagram showing fast-slow dynamics in phase space: $x$ represents a fast variable and $y_{1, 2}$ slow variables. Most of the time the dynamics remains on a strongly attracting "slow manifold". Trajectories of the system may hit "folds" where the attractor for the fast system loses stability leading to a rapid and possibly large reconfiguration (eg from $A$ to $B$) of the balance of climate variables. In reality there may be multiscale processes within the "slow manifold", and which variables are fast may change in different parts of phase space.

On the other hand, there has been steady progress in understanding the behaviours of forced nonlinear dynamical systems. If these two hands can come together, there may be very much to gain in terms of future understanding of climate variability, in particular for passage through non-stationary changes in forcing that lead to tipping points. Given the urgency to improve forecasts of the effects of anthropogenic forcing on variability, we suggest this should have a high research priority.

## Declaration of Competing Interest

The authors declare that they have no known competing financial interests or personal relationships that could have appeared to influence the work reported in this paper.

*Acknowledgement*

We thank the Past Earth Network (EPSRC grant EP/M008363/1) and ReCoVER (EPSRC grant EP/M008495/1) for supporting the workshop in 2017 where the idea of the paper was first discussed. This paper is TiPES contribution #22: this project has received funding from the European Union's Horizon 2020 research and innovation programme under grant agreement No 820970.